\documentclass[reprint,    
nofootinbib,  
aps,
physrev,
amsmath,
amssymb,
superscriptaddress,
]{revtex4-2}

\usepackage{graphicx} 
\usepackage{dcolumn}  
\usepackage{bm}       
\usepackage{float}

\usepackage[
colorlinks=true,
citecolor=blue
]{hyperref}

\usepackage{lipsum}
\usepackage{hepunits}
\usepackage{slashed}
\usepackage{ulem}

\graphicspath{{img/}} 

\usepackage{amsmath}
\usepackage{amssymb}
\usepackage{mathtools}
\usepackage{graphicx}
\usepackage{color}
\usepackage{comment}

\usepackage{booktabs}

\begin{document}

\title{
Dark Matter in the High-Scale Seesaw Leptogenesis Paradigm
}

\vspace{0.3cm}
\date{October 2024}

\author{Juan Herrero-Garcia}
\email{juanherrero@ific.uv.es}

\affiliation{Departament de Física Teòrica, Universitat de València, 46100 Burjassot, Spain}

\affiliation{Instituto de Física Corpuscular (CSIC-Universitat de València),
Parc Científic UV, C/Catedrático José Beltrán, 2, E-46980 Paterna, Spain}

\author{Giacomo Landini}
\email{giacomo.landini@ific.uv.es}

\affiliation{Instituto de Física Corpuscular (CSIC-Universitat de València),
Parc Científic UV, C/Catedrático José Beltrán, 2, E-46980 Paterna, Spain}

\author{Tsutomu T. Yanagida}
\email{tsutomu.tyanagida@sjtu.edu.cn}

\affiliation{Kavli Institute for the Physics and Mathematics of Universe (WPI), \\
University of Tokyo, Kashiwa 277-8583, Japan}

\begin{abstract}

The seesaw mechanism with three heavy Majorana right-handed neutrinos provides an elegant explanation for neutrino masses and, combined with leptogenesis, can generate the baryon asymmetry of the universe (BAU). Naturally embedded in a Grand Unified Theory, this framework stands as one of the best-motivated extensions beyond the Standard Model, but it is very difficult to test it. Moreover, it does not account for dark matter (DM). In this paper, we propose a minimal extension that introduces a dark sector with a singlet Majorana fermion (as the DM candidate) and a complex scalar singlet. The heavy right-handed neutrinos serve another role beyond generating neutrino masses and the BAU: producing the cold DM density through their decays. Interestingly, the model also predicts a subdominant DM component from late scalar decays, which in some cases may be hot or warm at the onset of structure formation, as well as an equal number of non-thermal neutrinos. These components leave distinct signatures in various cosmological observables. Furthermore, electromagnetic energy injection from scalar decays alter predictions from Big Bang Nucleosynthesis and induce spectral distortions in the Cosmic Microwave Background black-body spectrum. In this context, upcoming experiments, such as the Primordial Inflation Explorer (PIXIE), could probe the mechanism of neutrino mass generation.

\end{abstract}


\maketitle

\section{Introduction} \label{sec:intro}

The Standard Model (SM) lacks an explanation for the origin of neutrino masses ($m_\nu$), the baryon asymmetry of the universe (BAU), and the nature of dark matter (DM). One of the simplest extensions to generate light neutrino masses is the seesaw mechanism, where new heavy right-handed Majorana fermions $N_i$ ($i=1,2,3$, with masses $m_{N_i}$) with Yukawa couplings with the SM Higgs doublet $H$ and the lepton doublets $L_\alpha$ ($\alpha=e,\mu,\tau$), $\bar L y_\nu N \tilde H$, are added to the SM Lagrangian \cite{Minkowski:1977sc,Yanagida:1979as,Gell-Mann:1979vob,Yanagida:1979gs,Glashow:1979nm}.\footnote{It was first termed the ``seesaw mechanism" by one of the authors, T. T. Y., in a 1981 conference~\cite{Seesaw}.} This scenario may also be naturally embedded in Grand Unified Theories (GUTs). In particular, the right-handed neutrinos are needed to complete the fundamental representation of SO(10)~\cite{Gell-Mann:1979vob,Glashow:1979nm,Barbieri:1979ag} (see also Refs.~\cite{Wilczek:1979hh,Witten:1979nr,Weinberg:1979sa,PhysRevLett.44.912,Schechter:1980gr,Yanagida:1980xy} for further developments of the seesaw).\footnote{For a review of neutrino mass mechanisms, see Ref.~\cite{Cai:2017jrq}.}  In addition, the CP-violating out-of-equilibrium decays of the heavy right-handed neutrinos generate a lepton asymmetry which later on is transformed into a baryon one via sphaleron processes, fulfilling all Sakharov conditions \cite{Sakharov:1967dj}, in a process known as thermal leptogenesis \cite{Fukugita:1986hr,Buchmuller:2005eh,Davidson:2008bu}, as long as $m_N > 10^9$ GeV~\cite{Davidson:2002qv}.\footnote{It assumes a hierarchical spectrum of thermally produced right-handed neutrinos
without flavor effects.} A natural question, therefore, arises: \\

\emph{i) What is the simplest extension of this scenario that incorporates DM, ideally in a way that directly links it to the generation of $m_\nu$ and the BAU?\\}

Moreover, an equally relevant matter presents itself:\\

\emph{ii) Can the dark sector be used to test the otherwise untestable high-scale seesaw leptogenesis paradigm?\\}

Several works in the literature have addressed somewhat similar questions, one of the most popular proposals being the low-scale $\nu$MSM \cite{Akhmedov:1998qx,Shaposhnikov:2006xi,Klari__2021} (See also Refs.~\cite{Kitano:2004sv,Cosme:2005sb,An:2009vq,Hall:2010jx,Falkowski:2011xh,Chun:2011cc,Feng:2012jn,Unwin:2014poa,DiBari:2016guw,Escudero:2016tzx,Escudero:2016ksa,Gonzalez-Macias:2016vxy,Ballesteros_2017,Falkowski:2017uya,Biswas:2018sib,Caputo:2018zky,Chianese:2019epo,Bandyopadhyay_2020,Coy_2021,Datta:2021elq,Bhattacharya:2021jli,Sopov:2022bog,Coito:2022kif,Berbig:2023yyy,Li_2023,barman2023probingfreezeindarkmatter,Herrero-Garcia:2023lhv,Berbig:2024uwm}). In this work, we explore a minimal dark sector consisting of a Majorana singlet fermion $\chi$, with mass $m_\chi$, and a complex scalar singlet $S$, with mass $m_S$. In principle both $S$ and $\chi$ could be DM. However, with just interactions involving SM fields, it becomes challenging to generate the correct relic abundance for the fermion $\chi$. Additionally, scalar singlet DM via the Higgs portal, $\lambda_{HS} |S|^2 |H|^2$, faces significant constraints, even when it constitutes only a subdominant component, provided that the local energy density scales with the global one.\footnote{It is only allowed for $m_S=m_h/2$ GeV, where $m_h$ is the Higgs boson mass (i.e. at the resonance), or for $m_S\gtrsim 500$ GeV \cite{Cline_2013,GAMBIT:2017gge}.} Interestingly, thanks to the presence of heavy right-handed neutrinos, a new portal interaction with the SM, $y_S N \chi S$, is allowed. The latter changes  drastically the picture. In particular, this interaction enables two distinct mechanisms for producing sufficient DM in the early universe. Interestingly, this scenario leads to distinctive signatures that make it testable in future experiments.

In the following, for definiteness we consider $m_\chi < m_S$, so that the fermion is the DM candidate. For $m_S> m_N>10^9$ GeV, $S$ would have a too large abundance after freeze-out~\cite{Griest:1989wd}. Moreover, if this is the case, the DM $\chi$ cannot be experimentally detected as $S$ would need to decay before Big Bang Nucleosynthesis (BBN). However, if $m_S<m_N$, we have a hope to test the scenario, as we show in this paper. Thanks to the new portal interaction, the DM $\chi$ production occurs in a two-fold way: \emph{i)} via the freeze-in mechanism \cite{Hall:2009bx,Bernal_2017} through early decays, $N \rightarrow \chi+ S^\dagger$; and \emph{ii)} via late decays of the scalar into $\chi$ and active neutrinos through mixing, $S^\dagger \rightarrow \chi \nu$. The former decay generates a cold population of $\chi$, while the latter one gives rise to another subdominant $\chi$-component which in some cases could be hot/warm, and to dark radiation (non-thermal neutrinos $\nu$), both with equal number densities. Therefore, in order to obey the stringent limits on the latter, the $S$ population needs to be reduced before it decays. This is naturally achieved thanks to the unavoidable presence of the Higgs portal interaction (which also generates a thermal population of $S$). The mechanism is the following: the $S$ population from $N$ decays enters thermal equilibrium, in which $S^\dagger  S\rightarrow H^\dagger H$ annihilations further reduce it until they freeze-out; subsequently, the scalar $S$ has late decays into DM.\footnote{Late-decaying DM has also been proposed as a solution to the $H_0$ tension \cite{Vattis_2019,DiValentino:2021izs,Sch_neberg_2022}.} The production of $\chi$ from $S$ decays is an example of the super-WIMP mechanism \cite{Feng_2003,Feng_2004,Hall:2009bx,Cheung:2010gj,Garny:2018ali}.

In this work, we focus on the simultaneous generation of the BAU via high-scale thermal leptogenesis, and consider the thermal freeze-out of the scalar $S$, which is the most natural regime, i.e. having order one Higgs portal interactions. Furthermore, we embed the scenario in the well-motivated gauged B-L extension of the SM, which in turn can be straightforwardly integrated in SO(10). We also take into account bounds on dark radiation~\cite{Planck:2018vyg}, warm DM limits~\cite{Tan:2024cek}, as well as constraints from BBN~\cite{Cyburt:2015mya,Hambye_2022} and deviations of the Cosmic Microwave Background (CMB) energy spectrum from a black body (aka spectral distortions)~\cite{Acharya:2020gfh} due to electromagnetic energy injection from scalar decays. Gravitational waves may also be used to test the high-scale seesaw leptogenesis paradigm~\cite{Dror_2020}. Although similar scenarios and constraints have been considered in the literature~\cite{Falkowski:2011xh,Falkowski:2017uya,Becker:2018rve,Bandyopadhyay_2020,Li_2023,barman2023probingfreezeindarkmatter}, a unified and coherent proposal that incorporates all the essential elements is still lacking.

The remainder of this paper is organised as follows. In Sec.~\ref{sec:SS} we review the framework, which is an extension of the seesaw scenario embedded in a B-L gauge symmetry. In Sec.~\ref{sec:pheno} we analyse the different observables and constraints, and in Sec.~\ref{sec:results} we show the results of the numerical analysis. Finally we give our conclusions in Sec.~\ref{sec:conc}.

\section{The extended seesaw framework} \label{sec:SS}

\begin{table}[tbp!]
  \begin{center}
    \begin{tabular}{|c|c|c|c|}
    \hline
      {\bf Particle} & {\bf Notation}&   $U(1)_{\rm B-L}$ & $P_{M}$  \\ \hline
      Right-handed neutrinos & $N_i$ & -1  & - \\
      SM Higgs doublet &$H$& 0  & + \\
      Scalar singlet &$\sigma$ & 2  & + \\ \hline
      Dark Majorana fermion &$\chi$ & 0  & + \\
      Dark scalar singlet &$S$ & +1  & - \\
      \hline 
    \end{tabular}
  \end{center}
  \caption{\label{tab:Model}
   Charges under $U(1)_{\rm B-L}$ (third column) and the remnant matter parity $P_M=(-1)^{B-L}$ (fourth column) of the particles of the model. Standard Model leptons have $B-L=-1$, while quarks have $B-L=1/3$.}
\end{table}

We assume that the gauge $U(1)_{\rm  B-L}$ is preserved at very high energies. Therefore, anomaly cancellation requires three right-handed neutrinos $N_i$ ($i=1,2,3$). Their Majorana masses, $m_{N_1}\leq m_{N_2}\leq m_{N_3}$, are generated after the B-L symmetry is spontaneously broken at a very high scale by the VEV of the scalar singlet $\sigma$, $\langle \sigma \rangle=v_\sigma$. We consider that the associated gauge boson is even heavier so that it decouples from the low-energy spectrum. The spontaneous symmetry breaking (SSB) of B-L in two units generates at low energies an exact accidental remnant $Z_2$ symmetry, known as Matter Parity ($P_M$) (analogous to $R$-parity in Supersymmetry \cite{Martin:1992mq}), under which the lightest even (odd) particle, $\chi$ ($\nu$), is stable. In Table~\ref{tab:Model} we provide the particles and charges of the model under $U(1)_{\rm B-L}$ and $P_M$ in the third and fourth columns, respectively. 
\footnote{The model can be easily embedded in SO(10): $N$ in the fundamental ($\bf{16}$), $H$ in the $\bf{10}$, $\sigma$ in the $\bf{126}$, $\chi$ in the singlet ($\bf{1}$) and $S$ in the $\bf{16^*}$.} 

The new Lagrangian (density) can be written as $\mathcal{L}^{\rm new}_{\rm SM}=\mathcal{L}_{\rm SM} + \mathcal{L}_{\rm new}$, where $\mathcal{L}_{\rm SM}$ is the SM Lagrangian and the relevant new Yukawa interactions and fermion masses in $\mathcal{L}_{\rm new}$ are given by
\begin{equation}
-\mathcal{L}_{\rm new}\supset   \overline{L} \tilde H y_{\nu} N + \sigma \overline{N^c} y_\sigma  N+ 
S\overline{N^c}y_S \chi + \frac{1}{2} m_\chi\overline{\chi^c} \chi
+ {\rm H.c.}\,.
\end{equation}
Here $y_{\nu}$ is a $3 \times 3$ complex matrix and $y_\sigma$ is a $3 \times 3$ complex symmetric matrix which can be taken to be real and diagonal without loss of generality. Similarly, $y_S$ is a complex vector of Yukawa couplings. The relevant new terms in the scalar potential are:
\begin{equation} \label{eq:VSS}
-\mathcal{L}_{\rm new} \supset \mathcal{V}_{\rm new}\supset  \mu^2_S |S|^2 +  m^2_\sigma |\sigma|^2 +\lambda_{HS} |H|^2|S|^2\,.
\end{equation}
We take $\mu^2_S>0$ so that $S$ does not develop a Vacuum Expectation Value (VEV), whereas we consider $m^2_\sigma<0$. The electroweak (EW) symmetry is broken spontaneously when the SM Higgs doublet develops a VEV, $\langle H \rangle_0=v/\sqrt{2}$, with $v=246$ GeV.\footnote{A trilinear term, $\mu_{\sigma} \sigma^* S^2+ {\rm H.c.}$, which splits the real and imaginary parts of $S$, is also allowed, and generated radiatively, $\mu_{\sigma} \propto y_S^2 m_\chi/(4 \pi)^2$. It may wash-out any B-L asymmetry produced earlier on \cite{Harvey:1990qw}, so it should be small. Thus, $S$ is effectively a complex field, with mass $m_S$.
}

The Majorana mass matrix for the right-handed neutrinos, $m_N=y_\sigma v_\sigma$, is induced after the breaking of B-L. The Dirac mass matrix, $m_D=y_\nu v/\sqrt{2}$, is also generated after EW SSB. In the limit $m_N\gg m_D$, the mass of the active neutrinos is given by the seesaw formula~\cite{Minkowski:1977sc,Yanagida:1979as,Gell-Mann:1979vob,Glashow:1979nm,Yanagida:1980xy, PhysRevLett.44.912}, $m_\nu \simeq  m_D\,m_N^{-1} m^T_D$. To simplify the analysis, in the following we consider that the interactions of the dark sector are dominantly with the lightest of the right-handed neutrinos, $N_1 \equiv N$, so that $m_N$ is the mass of the latter, and $y_S$ is just a number that can be taken real without loss of generality. Reproducing the light neutrino mass scale with perturbative Yukawa couplings $y_\nu$, and generating the BAU for a hierarchical spectrum of right-handed neutrinos via thermal leptogenesis, requires $10^9\,{\rm  GeV}\lesssim m_N \lesssim 10^{15}\,{\rm  GeV}$~\cite{Davidson:2002qv}. For a detailed discussion of a similar two-sector leptogenesis for asymmetric DM, see Ref.~\cite{Falkowski:2011xh}, where it is demonstrated that for branching ratios ${\rm Br}_{\rm L} \gg {\rm Br}_\chi$ as in our case, the Davidson-Ibarra bound~\cite{Davidson:2002qv}, $m_N>10^9$ GeV, holds.

\section{Phenomenological constraints} \label{sec:pheno}

\subsection{Dark Matter relic abundance}
We are interested in the freeze-in production of $\chi$ from the decays of $N$. We assume that the reheating temperature fulfills $T_R>m_N$. Any pre-existing initial abundance of $\chi$ will have been depleted away by inflation. The condition to avoid that $\chi$ thermalises with the SM bath is\footnote{For such values of $y_S$, the production of $\chi$ from $2$-to-$2$ processes, $S^\dagger S\rightarrow \chi\chi$, mediated by the exchange of a heavy neutrino $N$, is suppressed.} \cite{Li_2023}
\begin{equation} \label{eq:fi}
\left(\frac{y_S}{10^{-2}} \right)^2 <\left(\frac{m_N}{3.5\cdot 10^{14}\,{\rm GeV}} \right)\,\left(\frac{g_*(T^N_{\rm d})}{g^{\rm SM}_*} \right)^{1/2}\,,
\end{equation}
where $g_*(T^N_{\rm d})$ ($g^{\rm SM}_*=106.75$) is the number of relativistic degrees of freedom when $N$ decays (of the SM particle content). Note that the Yukawa coupling may be much larger than in the usual freeze-in scenarios. The scalar $S$ has Higgs portal interactions, see Eq.~\eqref{eq:VSS}, which keep it in thermal equilibrium for $\lambda_{HS} \gtrsim 10^{-7}$ \cite{Enqvist:2014zqa}. After freeze-out, $S$ decays into the DM $\chi$ and active neutrinos, with decay rate \footnote{The three-body scalar decays $S^\dagger \rightarrow  \chi \nu h$ are subdominant for $m_S\lesssim$ TeV.}
\begin{equation} \label{eq:decS}
\Gamma_{S^\dagger \rightarrow \chi+\nu} \simeq \frac{y_S^2 m_S}{32 \pi}\,\left(\frac{m_\nu}{ m_N}\right)\left(1-\frac{m^2_\chi}{m^2_S}\right)^{2}\,.
\end{equation}
The lifetime of $S$, $\tau_S$, is always shorter than the age of the universe. Therefore, the relic abundance of $\chi$ today is given by the sum of both contributions:
\begin{equation}
\Omega_\chi h^2 = \Omega_N h^2 + \Omega_S \,\frac{E_\chi (|\vec p_0|)}{m_S}\,h^2\, ,
\end{equation}
where $h$ is the reduced Hubble constant and $E_\chi (|\vec p_0|)$ is the energy of $\chi$ at present times (redshifted from the time when $S$ decays), which can be computed from $S$ decays. The cold population generated by the decays of $N$ reads \cite{Hall:2009bx,Falkowski:2017uya,Li_2023} \footnote{It has been divided by 2 compared to Eq.~3.20 in Ref.~\cite{Li_2023} because the $S$ population is reduced by annihilations before decaying.}
\begin{equation} \label{eq:Omega}
 \frac{\Omega_N h^2}{\Omega_{\rm Pl} h^2} \simeq  \left(\frac{y_S}{10^{-5}} \right)^2\, \left(\frac{m_\chi}{{\rm GeV}}\right)\,\left(\frac{3.2\cdot 10^{13}\,{\rm GeV}}{m_N}\right)\,\left(\frac{g^{\rm SM}_*}{g_*(T^N_{\rm d})}\right)^{3/2}\,,
\end{equation}
where $\Omega_{\rm Pl} h^2= 0.120 \pm 0.001$ is the best-fit CDM value of the Planck satellite~\cite{Planck:2018vyg}. Note that the dominant contribution comes from the regime in which $N$ is in thermal equilibrium. The second component, which can be hot, warm or cold depending on the values of $m_\chi$ and $\tau_S$, is produced by the freeze-out of $S$ and subsequent decays into $\chi$. In the limit $m_S\gg m_h$, it reads
\begin{equation}
 \frac{\Omega_S h^2}{\Omega_{\rm Pl} h^2} \simeq\, \left(\frac{m_S}{{\rm TeV}} \right)^2\, \left(\frac{0.3}{\lambda_{HS}} \right)^2\,.
\end{equation}
In the numerical scan we use the full expression provided in Refs.~\cite{Cline:2012hg,Cline_2013}. A neat prediction of the scenario is that a hot neutrino contribution (i.e., dark radiation) from $S$ decays, with the same number density as the $\chi$ component, is also generated. Therefore, there could be a mixture of cold/warm/hot DM, and dark radiation, which is subject to constraints discussed below. For future convenience, we define the fraction of would-be $S$ DM if it was stable as $f_S \equiv \Omega_S/\Omega_{\rm Pl}$.

If the cold component from $N$ decays dominates, so that $\Omega_\chi h^2 \simeq \Omega_N h^2$, we can rewrite the lifetime of $S$ in terms of the total relic abundance as
\begin{equation}
\tau_S \simeq 10^{11}\,  {\rm s} \left(\frac{\Omega_{\rm Pl} h^2}{\Omega_\chi h^2}\right) \left(\frac{m_\chi}{m_S}\right) \lesssim 10^{11}  \,{\rm s}\,,
\end{equation}
where we have neglected the phase-space factor in Eq.~\eqref{eq:decS} and we have used the atmospheric neutrino mass scale, $m_\nu=0.05$ eV (See Ref.~\cite{esteban2024nufit60updatedglobalanalysis} for a very recent fit of neutrino oscillations). Curiously, in this regime the lifetime of $S$ is independent of the value of $m_N$. For any value of $m_\chi/m_S$, the decays are peaked before matter-radiation equality. A smaller ratio $m_\chi/m_S = 10\,{\rm keV}/(10\,{\rm TeV})=10^{-9}$ reduces the lifetime to $100$ s, around BBN. For $m_S<m_h/2$, there are important constraints from Higgs boson invisible decays at colliders, controlled by $\lambda_{HS}$. Therefore, we take $m_S \gtrsim m_h/2$ in the numerical analysis. Finally, direct detection signals of $\chi$, mediated by the Higgs portal, are very suppressed \cite{Gonzalez-Macias:2016vxy,Escudero:2016ksa,Herrero-Garcia:2018koq}. 

Bounds on warm dark matter (WDM) arise from its impact on structure formation at the smaller scales. Specifically, at the time of galaxy formation, DM cannot be too relativistic. Lyman-alpha observations of the 21cm Hydrogen line provide the strongest upper limits on the free-streaming length~\cite{Irsic:2017ixq,Boyarsky:2008xj,Decant:2021mhj}. For a given production mechanism, these constraints translate into lower bounds on the DM mass. In our scenario, for $m_\chi \gtrsim$ keV the $\chi$ population from $N$ early decays obeys these constraints \cite{Falkowski:2017uya}. The one originated from $S$ late decays, however, if it constitutes the total DM density, should fulfill \cite{Cirelli:2024ssz}: 
\begin{equation}
	m_\chi \gtrsim 1.9\,{\rm keV} \left\langle \frac{p}{T} \right\rangle_{\rm d} \,\left(\frac{g^{\rm SM}_*}{g_*(T^S_{\rm d})} \right)^{1/3}\,,
\end{equation}
where $g_*(T^S_{\rm d})$ is the number of relativistic degrees of freedom when $S$ decays and an average of the momentum-to-temperature ratio $\langle p/T \rangle_d$, instead of the full DM velocity distribution, has been employed. For a mixed cold plus warm scenario, if the fraction of WDM is below $60\%$, WDM masses as low as 5 keV are allowed \cite{Boyarsky:2008xj}, see also Ref.~\cite{Tan:2024cek}. For the case in which $\chi$ is hot, it contributes to dark radiation. In our analysis, we impose the limits from Fig.~3 of Ref.~\cite{Tan:2024cek}, rescaled by the momentum of the WDM particle. These are always satisfied for the small $\chi$ component stemming from $S$ decays.

\subsection{Limits from Big Bang Nucleosynthesis and the Cosmic Microwave Background}

There are several constraints arising from BBN and CMB observations. On the one hand, if the scalar decays occur before the time of recombination (the formation of nuclei), the free-streaming neutrinos (and also $\chi$ if it is relativistic at that time) produced from these decays contribute to dark radiation, leading to constraints from both BBN and CMB data. These limits are typically expressed as upper bounds on additional contributions to the effective number of neutrinos, $\Delta N_{\rm eff}$. The one-tailed Planck TT,TE,EE$+$lowE$+$lensing$+$BAO constraint reads $\Delta N_{\rm eff}< 0.30$ at $95\%\,{\rm CL}$ \cite{Planck:2018vyg}. In terms of the lifetime and fraction of $S$, the bound becomes approximately $\tau_S f_S^2 \lesssim 5 \cdot 10^9\,{\rm s}$~\cite{Hambye_2022}. The $\Delta N_{\rm eff}$ limit is expected to improve to a precision of $ \sigma(N_{\text{eff}}) \approx 0.05$ with the Simons Observatory~\cite{SimonsObservatory:2018koc}, and even to reach values as low as $\sigma(N_{\text{eff}}) \approx 0.014$ with EUCLID~\cite{EUCLID:2011zbd} and future generation experiments like CMB-HD~\cite{CMB-HD:2022bsz}.

\begin{figure*}[t!]
    \centering
	\includegraphics[width=0.46\textwidth]{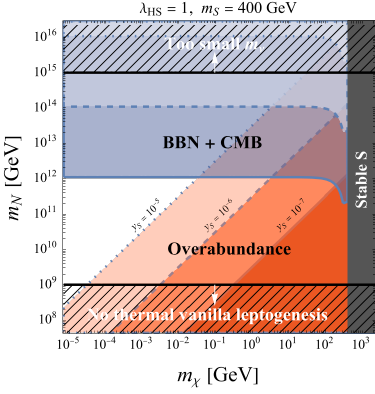}
    ~~~\includegraphics[width=0.51\textwidth]{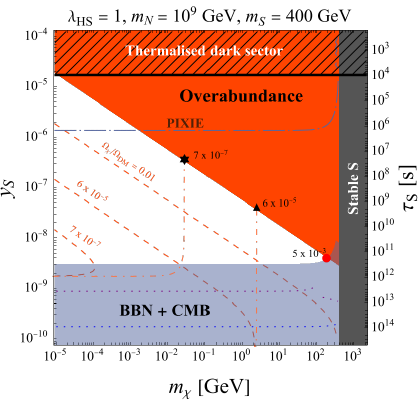}
    \caption{
    \emph{Left)} Plane of sterile neutrino and DM masses. The reddish (blueish) regions are excluded by overabundance (CMB and BBN limits); these are shown for $y_S=10^{-7}, 10^{-6},10^{-5}$ in solid, dashed, dotted, represented with a gradient from darker to lighter colors.
    \emph{Right)} Plane of the DM Yukawa coupling with sterile neutrinos (lifetime of $S$ on the right axis) and DM mass, for $m_N=10^9$ GeV. See text for further details.} \label{fig:mchi}	
\end{figure*}

On the other hand, if the scalar $S$ is massive enough the high-energy neutrinos produced from its decay radiate $W$ and $Z$ bosons (EW showering) that give rise to electromagnetic energy injection. Using this fact, in Refs.~\cite{Kanzaki:2007pd,Poulin:2016anj,Chluba:2019kpb,Bauer:2020jay,Chluba_2020,Acharya:2020gfh,Hambye_2022} constraints on long-lived particles decaying into neutrinos have been obtained from the observation of the CMB black-body spectrum, the measurement of the CMB anisotropies and the good agreement between theory and observations of the abundances of
light elements (modified by photo-disintegration processes induced by non-thermal photons produced from the EW showers). In particular, Ref.~\cite{Acharya:2020gfh} finds that for lifetimes in the $10^6-10^{10}$ s ballpark and neutrino energies above $\sim 100$ GeV, bounds from COBE-FIRAS data \cite{Mather:1993ij,Fixsen:1996nj} on the CMB black-body spectrum set much stronger limits on $N_{\rm eff}$ than Planck measurements of CMB anisotropies. CMB spectral distortions are deviations from a black-body spectrum. These are induced by EM energy injection (from $S$ decays into $\nu$ and the subsequent EW showering) which modifies the CMB spectrum when (Compton) scatterings and number-changing processes (such as double Compton scatterings) are not efficient enough to re-establish a black-body spectrum. On the other hand, if $S$ decays at later times, the resulting energy injection modifies the evolution of the free electron fraction and recombination. Therefore, measurements of CMB anisotropies provide the strongest constraints for lifetimes $\tau_S\gtrsim 10^{11}$s.

The current leading constraints for lifetimes in the range $10^5-10^{10}$s come from BBN observations~\cite{Li_2021,Hambye_2022}, which exclude masses of the decaying particle above $\sim 100$ GeV for lifetimes $\gtrsim 10^5$ s if the decaying particle constitutes the whole DM energy density. However, future prospects of spectral distortion measurements of the CMB black body spectrum may improve the current bounds by several orders of magnitude with the ``The Primordial Inflation Explorer (PIXIE)"~\cite{Kogut:2011xw}, see also Fig. 13 in Ref.~\cite{Acharya:2020gfh}. In our numerical analysis, we impose the limits presented in Figs. 2 and 3 of Ref.~\cite{Hambye_2022} rescaled by a factor of two to take into account the different number of neutrinos in the final state, together with future prospects~\cite{Kogut:2011xw}. Finally, the high-energy neutrino flux~\cite{Feldstein:2013kka,Palomares-Ruiz:2007egs,Bell:2010fk,Esmaili:2012us,Poulin:2016nat,Arguelles:2019ouk, Palomares-Ruiz:2020ytu,Bandyopadhyay_2020}, that could potentially be detected at neutrino detectors~\cite{Super-Kamiokande:2002weg,Hyper-Kamiokande:2018ofw,Aartsen_2013,IceCube:2018tkk,ANTARES:2015vis,KM3Net:2016zxf,KM3NeT:2018wnd}, is very suppressed.

\section{Numerical Analysis} \label{sec:results}

There are only a few relevant independent parameters in the model: $m_N$, $m_\chi$, $m_S$ and $\lambda_{HS}$. Neutrino masses fix $y_\nu$. The DM relic abundance, if it is mainly composed of the cold component from $N$ decays, essentially fixes $y_S$. In Fig.~\ref{fig:mchi} we plot the planes $m_N$ ($y_S$) versus $m_\chi$ in the left (right) panel. We fix $m_S=400$ GeV and $\lambda_{HS}=1$, so that that $f_S=0.01$ (similar results are obtained for larger $m_S\simeq \mathcal{O} (1)$ TeV). In all the plane of the left plot, the freeze-in condition given in Eq.~\eqref{eq:fi} is satisfied. We highlight as hatched the region where vanilla thermal leptogenesis is not possible and the generated neutrino masses are too small for perturbative Yukawas. The grey region indicates where $S$ is stable (excluded by direct/indirect detection), which occurs only when the decays of $S$ are kinematically forbidden (for $m_\chi > m_S$). In the reddish regions, the total DM abundance is overproduced, while the blueish areas are ruled out by BBN and CMB constraints~\cite{Hambye_2022}. The latter two are complementary: the larger the Yukawa, the larger (smaller) the region excluded by overabundance (BBN+CMB). Their intersection provides an upper limit for the DM mass, $m_\chi \lesssim 200$ GeV. 

As can be observed, there is a significant portion of the parameter space where all three SM shortcomings ($m_\nu$, BAU, DM) can be successfully addressed. Also, heavy sterile neutrino masses, typically untestable, are excluded in this scenario by overabundance and/or by cosmology. In particular, $m_N<10^{12}$ GeV for small Yukawas, $y_S=10^{-7}$ (corresponding to $m_\chi \lesssim 200$ GeV). For larger $y_S$, a lower bound on $m_N$ stronger than $10^9$ GeV appears. For $y_S \rightarrow 0, \lambda_{HS} \rightarrow 0$, the standard seesaw leptogenesis paradigm is recovered.

In the right plot of Fig.~\ref{fig:mchi}, we fix $m_N=10^9$ GeV. In the hatched region the freeze-in condition is not satisfied. The red, blue and gray regions have the same meaning as in the left figure. We show the $\Omega_\chi/\Omega_{\rm Pl} = 1$ contour with a solid line, as well as the contours of total relic abundance: $\Omega_\chi/\Omega_{\rm Pl} = 0.01, 6 \times 10^{-5}, 7 \times 10^{-7}$, with dashed lines. We also present the current (future) $\Delta N_{\rm eff}<0.3$ ($\Delta N_{\rm eff}<0.05$) limit by Planck~\cite{Planck:2018vyg} (Simons Observatory~\cite{SimonsObservatory:2018koc}) as a blue (purple) dotted line. Note that the \emph{kink} in the $\Delta N_{\rm eff}$ constraints is due to the fact that $\chi$ does not contribute for large enough masses (to the right of the kink) because it is already non-relativistic at CMB times. The dot long-dashed line indicates a (conservative) projected sensitivity of PIXIE~\cite{Acharya:2020gfh,Kogut:2011xw}. The latter would be able to probe a large region of the parameter space. This shows that precise measurements of the CMB spectral distortions could provide a powerful tool to test the high-scale seesaw leptogenesis scenario, in a way closely related to the existence of the dark sector.

In the relic abundance contours of the right panel of Fig.~\ref{fig:mchi}, there are two different regimes: \emph{i)} the top diagonal line with $\tau_S \propto m_\chi$, that is dominated by the freeze-in contribution, \emph{ii)} the right  (lower) part, where $S$ decays are significant, and $\chi$ is cold (hot). Lifetimes of $S$ larger than the age of the universe (not shown in the plots) are ruled-out by direct/indirect searches irrespective of whether $S$ is a subdominant component of DM. The star (triangle) shows the value of lifetime below which (and also the DM mass to the left of which) the (small) $\chi$ component from $S$ decays is hot (warm), $|\vec p|\gtrsim 10 \,m_\chi$ ($|\vec p|\gtrsim m_\chi$), at structure formation timescales. In these points, the fractions of hot and warm DM from $S$ decays (written next to them) are maximum. The red dot highlights the maximum allowed fraction of (cold) $\chi$ from $S$ decays. The dot-dashed lines starting from the star/triangle are contours of constant $\chi$ fraction originated from $S$ decays. For large (low) enough values of $\tau_S$ ($m_\chi$), the freeze-in contribution is irrelevant, as can be seen by the contours of constant $\chi$ fraction originated from $S$ decays asymptotically approaching the total relic abundance contours, i.e. $\Omega_\chi \rightarrow \Omega_S$.

\section{Conclusions} \label{sec:conc}
The new SM should provide an explanation for neutrino masses, the baryon asymmetry of the universe, and dark matter. The high-scale seesaw leptogenesis paradigm, embedded within a B-L gauge symmetry, is one of the best-motivated extensions of the SM to address neutrino masses and the matter-antimatter asymmetry, but it does not offer a solution to the DM conundrum. 


While a simple option for the DM puzzle would be a massive stable fermion with only gravitational interactions, this would be undetectable (``nightmare" DM) and extremely heavy.  In our paper, however, we highlight that this fermion \(\chi\), stable due to the $Z_2$ (exact) remnant symmetry of the ($B-L$) gauge symmetry, can couple to right-handed neutrinos (\(N\)) if we introduce a new scalar field, \(S\), which serves as a partner to \(N\).

The presence of $S$ and the portal with sterile neutrinos makes the physics of the DM $\chi$ very rich. The latter is primarily produced through the freeze-in mechanism, which in this case does not require very small Yukawa couplings as the right-handed neutrinos are very heavy, i.e., GUT scale. The abundance of the scalar $S$ is suppressed by large Higgs portal interactions. A subdominant component of $\chi$, which may be hot, warm or cold, is produced by the late decays of $S$, which in addition also generate a small component of non-thermal neutrinos in equal numbers. The latter subsequently also lead to electromagnetic energy injection which alter BBN and CMB predictions. 

The scenario presented here constitutes a simple extension of the SM, elegantly addressing some of its most significant shortcomings. Importantly, our work highlights the under-appreciated role of cosmological signatures in the seesaw paradigm, underscoring their importance as a promising avenue for future research and experimental validation. In particular, the small component of warm/hot DM, along with future measurements of CMB spectral distortions in PIXIE~\cite{Kogut:2011xw}, serve as potential probes of the proposal. 

\vspace{0.5 cm}

\section*{Acknowledgements}
We are thankful to Maximilian Berbig for a careful reading of the manuscript and his useful comments. We are grateful to the University of Tokyo and the Kavli Institute of Physics and Mathematics of the Universe (IPMU) 
for their warm hospitality, which made JHG and GL's research stay, during which this work was conducted, highly enjoyable. This project has received funding from the European Union’s Horizon Europe research and innovation programme under the Marie Skłodowska-Curie Staff Exchange  grant agreement No 101086085 – ASYMMETRY. JHG  is supported by \emph{Consolidación Investigadora Grant CNS2022-135592}, funded also by \emph{European Union NextGenerationEU/PRTR}. GL is supported by the Generalitat Valenciana APOSTD/2023 Grant No. CIAPOS/2022/193. This work is partially supported by the Spanish \emph{Agencia Estatal de Investigación} MICINN/AEI (10.13039/501100011033) grant PID2023-148162NB-C21 and the \emph{Severo Ochoa} project MCIU/AEI CEX2023-001292-S.
This work is also supported by the JSPS Grant-in-Aid for Scientific Research Grants No.\,24H02244, the National Natural Science
Foundation of China (12175134) and the World Premier International Research Center Initiative (WPI Initiative), MEXT, Japan.

\bibliography{main}

\begin{thebibliography}{105}%
\makeatletter
\providecommand \@ifxundefined [1]{%
 \@ifx{#1\undefined}
}%
\providecommand \@ifnum [1]{%
 \ifnum #1\expandafter \@firstoftwo
 \else \expandafter \@secondoftwo
 \fi
}%
\providecommand \@ifx [1]{%
 \ifx #1\expandafter \@firstoftwo
 \else \expandafter \@secondoftwo
 \fi
}%
\providecommand \natexlab [1]{#1}%
\providecommand \enquote  [1]{``#1''}%
\providecommand \bibnamefont  [1]{#1}%
\providecommand \bibfnamefont [1]{#1}%
\providecommand \citenamefont [1]{#1}%
\providecommand \href@noop [0]{\@secondoftwo}%
\providecommand \href [0]{\begingroup \@sanitize@url \@href}%
\providecommand \@href[1]{\@@startlink{#1}\@@href}%
\providecommand \@@href[1]{\endgroup#1\@@endlink}%
\providecommand \@sanitize@url [0]{\catcode `\\12\catcode `\$12\catcode
  `\&12\catcode `\#12\catcode `\^12\catcode `\_12\catcode `\%12\relax}%
\providecommand \@@startlink[1]{}%
\providecommand \@@endlink[0]{}%
\providecommand \url  [0]{\begingroup\@sanitize@url \@url }%
\providecommand \@url [1]{\endgroup\@href {#1}{\urlprefix }}%
\providecommand \urlprefix  [0]{URL }%
\providecommand \Eprint [0]{\href }%
\providecommand \doibase [0]{https://doi.org/}%
\providecommand \selectlanguage [0]{\@gobble}%
\providecommand \bibinfo  [0]{\@secondoftwo}%
\providecommand \bibfield  [0]{\@secondoftwo}%
\providecommand \translation [1]{[#1]}%
\providecommand \BibitemOpen [0]{}%
\providecommand \bibitemStop [0]{}%
\providecommand \bibitemNoStop [0]{.\EOS\space}%
\providecommand \EOS [0]{\spacefactor3000\relax}%
\providecommand \BibitemShut  [1]{\csname bibitem#1\endcsname}%
\let\auto@bib@innerbib\@empty
\bibitem [{\citenamefont {Minkowski}(1977)}]{Minkowski:1977sc}%
  \BibitemOpen
  \bibfield  {author} {\bibinfo {author} {\bibfnamefont {P.}~\bibnamefont
  {Minkowski}},\ }\bibfield  {title} {\bibinfo {title} {{$\mu \to e\gamma$ at a
  Rate of One Out of $10^{9}$ Muon Decays?}},\ }\href
  {https://doi.org/10.1016/0370-2693(77)90435-X} {\bibfield  {journal}
  {\bibinfo  {journal} {Phys. Lett. B}\ }\textbf {\bibinfo {volume} {67}},\
  \bibinfo {pages} {421} (\bibinfo {year} {1977})}\BibitemShut {NoStop}%
\bibitem [{\citenamefont {Yanagida}(1979{\natexlab{a}})}]{Yanagida:1979as}%
  \BibitemOpen
  \bibfield  {author} {\bibinfo {author} {\bibfnamefont {T.}~\bibnamefont
  {Yanagida}},\ }\bibfield  {title} {\bibinfo {title} {{Horizontal gauge
  symmetry and masses of neutrinos}},\ }\href@noop {} {\bibfield  {journal}
  {\bibinfo  {journal} {Conf. Proc. C}\ }\textbf {\bibinfo {volume}
  {7902131}},\ \bibinfo {pages} {95} (\bibinfo {year}
  {1979}{\natexlab{a}})}\BibitemShut {NoStop}%
\bibitem [{\citenamefont {Gell-Mann}\ \emph {et~al.}(1979)\citenamefont
  {Gell-Mann}, \citenamefont {Ramond},\ and\ \citenamefont
  {Slansky}}]{Gell-Mann:1979vob}%
  \BibitemOpen
  \bibfield  {author} {\bibinfo {author} {\bibfnamefont {M.}~\bibnamefont
  {Gell-Mann}}, \bibinfo {author} {\bibfnamefont {P.}~\bibnamefont {Ramond}},\
  and\ \bibinfo {author} {\bibfnamefont {R.}~\bibnamefont {Slansky}},\
  }\bibfield  {title} {\bibinfo {title} {{Complex Spinors and Unified
  Theories}},\ }\href@noop {} {\bibfield  {journal} {\bibinfo  {journal} {Conf.
  Proc. C}\ }\textbf {\bibinfo {volume} {790927}},\ \bibinfo {pages} {315}
  (\bibinfo {year} {1979})},\ \Eprint {https://arxiv.org/abs/1306.4669}
  {arXiv:1306.4669 [hep-th]} \BibitemShut {NoStop}%
\bibitem [{\citenamefont {Yanagida}(1979{\natexlab{b}})}]{Yanagida:1979gs}%
  \BibitemOpen
  \bibfield  {author} {\bibinfo {author} {\bibfnamefont {T.}~\bibnamefont
  {Yanagida}},\ }\bibfield  {title} {\bibinfo {title} {{Horizontal Symmetry and
  Mass of the Top Quark}},\ }\href {https://doi.org/10.1103/PhysRevD.20.2986}
  {\bibfield  {journal} {\bibinfo  {journal} {Phys. Rev. D}\ }\textbf {\bibinfo
  {volume} {20}},\ \bibinfo {pages} {2986} (\bibinfo {year}
  {1979}{\natexlab{b}})}\BibitemShut {NoStop}%
\bibitem [{\citenamefont {Glashow}(1980)}]{Glashow:1979nm}%
  \BibitemOpen
  \bibfield  {author} {\bibinfo {author} {\bibfnamefont {S.~L.}\ \bibnamefont
  {Glashow}},\ }\bibfield  {title} {\bibinfo {title} {{The Future of Elementary
  Particle Physics}},\ }\href {https://doi.org/10.1007/978-1-4684-7197-7_15}
  {\bibfield  {journal} {\bibinfo  {journal} {NATO Sci. Ser. B}\ }\textbf
  {\bibinfo {volume} {61}},\ \bibinfo {pages} {687} (\bibinfo {year}
  {1980})}\BibitemShut {NoStop}%
\bibitem [{See(1981)}]{Seesaw}%
  \BibitemOpen
  \bibfield  {title} {\bibinfo {title} {{1981 INS Symposium on Quark and Lepton
  Physics, Tokyo}},\ }\href@noop {} {\  (\bibinfo {year} {June 25-27,
  1981})}\BibitemShut {NoStop}%
\bibitem [{\citenamefont {Barbieri}\ \emph {et~al.}(1980)\citenamefont
  {Barbieri}, \citenamefont {Nanopoulos}, \citenamefont {Morchio},\ and\
  \citenamefont {Strocchi}}]{Barbieri:1979ag}%
  \BibitemOpen
  \bibfield  {author} {\bibinfo {author} {\bibfnamefont {R.}~\bibnamefont
  {Barbieri}}, \bibinfo {author} {\bibfnamefont {D.~V.}\ \bibnamefont
  {Nanopoulos}}, \bibinfo {author} {\bibfnamefont {G.}~\bibnamefont
  {Morchio}},\ and\ \bibinfo {author} {\bibfnamefont {F.}~\bibnamefont
  {Strocchi}},\ }\bibfield  {title} {\bibinfo {title} {{Neutrino Masses in
  Grand Unified Theories}},\ }\href
  {https://doi.org/10.1016/0370-2693(80)90058-1} {\bibfield  {journal}
  {\bibinfo  {journal} {Phys. Lett. B}\ }\textbf {\bibinfo {volume} {90}},\
  \bibinfo {pages} {91} (\bibinfo {year} {1980})}\BibitemShut {NoStop}%
\bibitem [{\citenamefont {Wilczek}(1979)}]{Wilczek:1979hh}%
  \BibitemOpen
  \bibfield  {author} {\bibinfo {author} {\bibfnamefont {F.}~\bibnamefont
  {Wilczek}},\ }\bibfield  {title} {\bibinfo {title} {{Unification of
  Fundamental Forces}},\ }\href@noop {} {\bibfield  {journal} {\bibinfo
  {journal} {eConf}\ }\textbf {\bibinfo {volume} {C790823}},\ \bibinfo {pages}
  {437} (\bibinfo {year} {1979})}\BibitemShut {NoStop}%
\bibitem [{\citenamefont {Witten}(1980)}]{Witten:1979nr}%
  \BibitemOpen
  \bibfield  {author} {\bibinfo {author} {\bibfnamefont {E.}~\bibnamefont
  {Witten}},\ }\bibfield  {title} {\bibinfo {title} {{Neutrino Masses in the
  Minimal O(10) Theory}},\ }\href
  {https://doi.org/10.1016/0370-2693(80)90666-8} {\bibfield  {journal}
  {\bibinfo  {journal} {Phys. Lett. B}\ }\textbf {\bibinfo {volume} {91}},\
  \bibinfo {pages} {81} (\bibinfo {year} {1980})}\BibitemShut {NoStop}%
\bibitem [{\citenamefont {Weinberg}(1979)}]{Weinberg:1979sa}%
  \BibitemOpen
  \bibfield  {author} {\bibinfo {author} {\bibfnamefont {S.}~\bibnamefont
  {Weinberg}},\ }\bibfield  {title} {\bibinfo {title} {{Baryon and Lepton
  Nonconserving Processes}},\ }\href
  {https://doi.org/10.1103/PhysRevLett.43.1566} {\bibfield  {journal} {\bibinfo
   {journal} {Phys. Rev. Lett.}\ }\textbf {\bibinfo {volume} {43}},\ \bibinfo
  {pages} {1566} (\bibinfo {year} {1979})}\BibitemShut {NoStop}%
\bibitem [{\citenamefont {Mohapatra}\ and\ \citenamefont
  {Senjanovi\ifmmode~\acute{c}\else \'{c}\fi{}}(1980)}]{PhysRevLett.44.912}%
  \BibitemOpen
  \bibfield  {author} {\bibinfo {author} {\bibfnamefont {R.~N.}\ \bibnamefont
  {Mohapatra}}\ and\ \bibinfo {author} {\bibfnamefont {G.}~\bibnamefont
  {Senjanovi\ifmmode~\acute{c}\else \'{c}\fi{}}},\ }\bibfield  {title}
  {\bibinfo {title} {Neutrino mass and spontaneous parity nonconservation},\
  }\href {https://doi.org/10.1103/PhysRevLett.44.912} {\bibfield  {journal}
  {\bibinfo  {journal} {Phys. Rev. Lett.}\ }\textbf {\bibinfo {volume} {44}},\
  \bibinfo {pages} {912} (\bibinfo {year} {1980})}\BibitemShut {NoStop}%
\bibitem [{\citenamefont {Schechter}\ and\ \citenamefont
  {Valle}(1980)}]{Schechter:1980gr}%
  \BibitemOpen
  \bibfield  {author} {\bibinfo {author} {\bibfnamefont {J.}~\bibnamefont
  {Schechter}}\ and\ \bibinfo {author} {\bibfnamefont {J.~W.~F.}\ \bibnamefont
  {Valle}},\ }\bibfield  {title} {\bibinfo {title} {{Neutrino Masses in SU(2) x
  U(1) Theories}},\ }\href {https://doi.org/10.1103/PhysRevD.22.2227}
  {\bibfield  {journal} {\bibinfo  {journal} {Phys. Rev. D}\ }\textbf {\bibinfo
  {volume} {22}},\ \bibinfo {pages} {2227} (\bibinfo {year}
  {1980})}\BibitemShut {NoStop}%
\bibitem [{\citenamefont {Yanagida}(1980)}]{Yanagida:1980xy}%
  \BibitemOpen
  \bibfield  {author} {\bibinfo {author} {\bibfnamefont {T.}~\bibnamefont
  {Yanagida}},\ }\bibfield  {title} {\bibinfo {title} {{Horizontal Symmetry and
  Masses of Neutrinos}},\ }\href {https://doi.org/10.1143/PTP.64.1103}
  {\bibfield  {journal} {\bibinfo  {journal} {Prog. Theor. Phys.}\ }\textbf
  {\bibinfo {volume} {64}},\ \bibinfo {pages} {1103} (\bibinfo {year}
  {1980})}\BibitemShut {NoStop}%
\bibitem [{\citenamefont {Cai}\ \emph {et~al.}(2017)\citenamefont {Cai},
  \citenamefont {Herrero-Garc\'\i{}a}, \citenamefont {Schmidt}, \citenamefont
  {Vicente},\ and\ \citenamefont {Volkas}}]{Cai:2017jrq}%
  \BibitemOpen
  \bibfield  {author} {\bibinfo {author} {\bibfnamefont {Y.}~\bibnamefont
  {Cai}}, \bibinfo {author} {\bibfnamefont {J.}~\bibnamefont
  {Herrero-Garc\'\i{}a}}, \bibinfo {author} {\bibfnamefont {M.~A.}\
  \bibnamefont {Schmidt}}, \bibinfo {author} {\bibfnamefont {A.}~\bibnamefont
  {Vicente}},\ and\ \bibinfo {author} {\bibfnamefont {R.~R.}\ \bibnamefont
  {Volkas}},\ }\bibfield  {title} {\bibinfo {title} {{From the trees to the
  forest: a review of radiative neutrino mass models}},\ }\href
  {https://doi.org/10.3389/fphy.2017.00063} {\bibfield  {journal} {\bibinfo
  {journal} {Front. in Phys.}\ }\textbf {\bibinfo {volume} {5}},\ \bibinfo
  {pages} {63} (\bibinfo {year} {2017})},\ \Eprint
  {https://arxiv.org/abs/1706.08524} {arXiv:1706.08524 [hep-ph]} \BibitemShut
  {NoStop}%
\bibitem [{\citenamefont {Sakharov}(1967)}]{Sakharov:1967dj}%
  \BibitemOpen
  \bibfield  {author} {\bibinfo {author} {\bibfnamefont {A.~D.}\ \bibnamefont
  {Sakharov}},\ }\bibfield  {title} {\bibinfo {title} {{Violation of CP
  Invariance, C asymmetry, and baryon asymmetry of the universe}},\ }\href
  {https://doi.org/10.1070/PU1991v034n05ABEH002497} {\bibfield  {journal}
  {\bibinfo  {journal} {Pisma Zh. Eksp. Teor. Fiz.}\ }\textbf {\bibinfo
  {volume} {5}},\ \bibinfo {pages} {32} (\bibinfo {year} {1967})}\BibitemShut
  {NoStop}%
\bibitem [{\citenamefont {Fukugita}\ and\ \citenamefont
  {Yanagida}(1986)}]{Fukugita:1986hr}%
  \BibitemOpen
  \bibfield  {author} {\bibinfo {author} {\bibfnamefont {M.}~\bibnamefont
  {Fukugita}}\ and\ \bibinfo {author} {\bibfnamefont {T.}~\bibnamefont
  {Yanagida}},\ }\bibfield  {title} {\bibinfo {title} {{Baryogenesis Without
  Grand Unification}},\ }\href {https://doi.org/10.1016/0370-2693(86)91126-3}
  {\bibfield  {journal} {\bibinfo  {journal} {Phys. Lett. B}\ }\textbf
  {\bibinfo {volume} {174}},\ \bibinfo {pages} {45} (\bibinfo {year}
  {1986})}\BibitemShut {NoStop}%
\bibitem [{\citenamefont {Buchmuller}\ \emph {et~al.}(2005)\citenamefont
  {Buchmuller}, \citenamefont {Peccei},\ and\ \citenamefont
  {Yanagida}}]{Buchmuller:2005eh}%
  \BibitemOpen
  \bibfield  {author} {\bibinfo {author} {\bibfnamefont {W.}~\bibnamefont
  {Buchmuller}}, \bibinfo {author} {\bibfnamefont {R.~D.}\ \bibnamefont
  {Peccei}},\ and\ \bibinfo {author} {\bibfnamefont {T.}~\bibnamefont
  {Yanagida}},\ }\bibfield  {title} {\bibinfo {title} {{Leptogenesis as the
  origin of matter}},\ }\href
  {https://doi.org/10.1146/annurev.nucl.55.090704.151558} {\bibfield  {journal}
  {\bibinfo  {journal} {Ann. Rev. Nucl. Part. Sci.}\ }\textbf {\bibinfo
  {volume} {55}},\ \bibinfo {pages} {311} (\bibinfo {year} {2005})},\ \Eprint
  {https://arxiv.org/abs/hep-ph/0502169} {arXiv:hep-ph/0502169} \BibitemShut
  {NoStop}%
\bibitem [{\citenamefont {Davidson}\ \emph {et~al.}(2008)\citenamefont
  {Davidson}, \citenamefont {Nardi},\ and\ \citenamefont
  {Nir}}]{Davidson:2008bu}%
  \BibitemOpen
  \bibfield  {author} {\bibinfo {author} {\bibfnamefont {S.}~\bibnamefont
  {Davidson}}, \bibinfo {author} {\bibfnamefont {E.}~\bibnamefont {Nardi}},\
  and\ \bibinfo {author} {\bibfnamefont {Y.}~\bibnamefont {Nir}},\ }\bibfield
  {title} {\bibinfo {title} {{Leptogenesis}},\ }\href
  {https://doi.org/10.1016/j.physrep.2008.06.002} {\bibfield  {journal}
  {\bibinfo  {journal} {Phys. Rept.}\ }\textbf {\bibinfo {volume} {466}},\
  \bibinfo {pages} {105} (\bibinfo {year} {2008})},\ \Eprint
  {https://arxiv.org/abs/0802.2962} {arXiv:0802.2962 [hep-ph]} \BibitemShut
  {NoStop}%
\bibitem [{\citenamefont {Davidson}\ and\ \citenamefont
  {Ibarra}(2002)}]{Davidson:2002qv}%
  \BibitemOpen
  \bibfield  {author} {\bibinfo {author} {\bibfnamefont {S.}~\bibnamefont
  {Davidson}}\ and\ \bibinfo {author} {\bibfnamefont {A.}~\bibnamefont
  {Ibarra}},\ }\bibfield  {title} {\bibinfo {title} {{A Lower bound on the
  right-handed neutrino mass from leptogenesis}},\ }\href
  {https://doi.org/10.1016/S0370-2693(02)01735-5} {\bibfield  {journal}
  {\bibinfo  {journal} {Phys. Lett. B}\ }\textbf {\bibinfo {volume} {535}},\
  \bibinfo {pages} {25} (\bibinfo {year} {2002})},\ \Eprint
  {https://arxiv.org/abs/hep-ph/0202239} {arXiv:hep-ph/0202239} \BibitemShut
  {NoStop}%
\bibitem [{\citenamefont {Akhmedov}\ \emph {et~al.}(1998)\citenamefont
  {Akhmedov}, \citenamefont {Rubakov},\ and\ \citenamefont
  {Smirnov}}]{Akhmedov:1998qx}%
  \BibitemOpen
  \bibfield  {author} {\bibinfo {author} {\bibfnamefont {E.~K.}\ \bibnamefont
  {Akhmedov}}, \bibinfo {author} {\bibfnamefont {V.~A.}\ \bibnamefont
  {Rubakov}},\ and\ \bibinfo {author} {\bibfnamefont {A.~Y.}\ \bibnamefont
  {Smirnov}},\ }\bibfield  {title} {\bibinfo {title} {{Baryogenesis via
  neutrino oscillations}},\ }\href
  {https://doi.org/10.1103/PhysRevLett.81.1359} {\bibfield  {journal} {\bibinfo
   {journal} {Phys. Rev. Lett.}\ }\textbf {\bibinfo {volume} {81}},\ \bibinfo
  {pages} {1359} (\bibinfo {year} {1998})},\ \Eprint
  {https://arxiv.org/abs/hep-ph/9803255} {arXiv:hep-ph/9803255} \BibitemShut
  {NoStop}%
\bibitem [{\citenamefont {Shaposhnikov}\ and\ \citenamefont
  {Tkachev}(2006)}]{Shaposhnikov:2006xi}%
  \BibitemOpen
  \bibfield  {author} {\bibinfo {author} {\bibfnamefont {M.}~\bibnamefont
  {Shaposhnikov}}\ and\ \bibinfo {author} {\bibfnamefont {I.}~\bibnamefont
  {Tkachev}},\ }\bibfield  {title} {\bibinfo {title} {{The nuMSM, inflation,
  and dark matter}},\ }\href {https://doi.org/10.1016/j.physletb.2006.06.063}
  {\bibfield  {journal} {\bibinfo  {journal} {Phys. Lett. B}\ }\textbf
  {\bibinfo {volume} {639}},\ \bibinfo {pages} {414} (\bibinfo {year}
  {2006})},\ \Eprint {https://arxiv.org/abs/hep-ph/0604236}
  {arXiv:hep-ph/0604236} \BibitemShut {NoStop}%
\bibitem [{\citenamefont {Klarić}\ \emph {et~al.}(2021)\citenamefont
  {Klarić}, \citenamefont {Shaposhnikov},\ and\ \citenamefont
  {Timiryasov}}]{Klari__2021}%
  \BibitemOpen
  \bibfield  {author} {\bibinfo {author} {\bibfnamefont {J.}~\bibnamefont
  {Klarić}}, \bibinfo {author} {\bibfnamefont {M.}~\bibnamefont
  {Shaposhnikov}},\ and\ \bibinfo {author} {\bibfnamefont {I.}~\bibnamefont
  {Timiryasov}},\ }\bibfield  {title} {\bibinfo {title} {Uniting low-scale
  leptogenesis mechanisms},\ }\bibfield  {journal} {\bibinfo  {journal}
  {Physical Review Letters}\ }\textbf {\bibinfo {volume} {127}},\ \href
  {https://doi.org/10.1103/physrevlett.127.111802}
  {10.1103/physrevlett.127.111802} (\bibinfo {year} {2021})\BibitemShut
  {NoStop}%
\bibitem [{\citenamefont {Kitano}\ and\ \citenamefont
  {Low}(2005)}]{Kitano:2004sv}%
  \BibitemOpen
  \bibfield  {author} {\bibinfo {author} {\bibfnamefont {R.}~\bibnamefont
  {Kitano}}\ and\ \bibinfo {author} {\bibfnamefont {I.}~\bibnamefont {Low}},\
  }\bibfield  {title} {\bibinfo {title} {{Dark matter from baryon asymmetry}},\
  }\href {https://doi.org/10.1103/PhysRevD.71.023510} {\bibfield  {journal}
  {\bibinfo  {journal} {Phys. Rev. D}\ }\textbf {\bibinfo {volume} {71}},\
  \bibinfo {pages} {023510} (\bibinfo {year} {2005})},\ \Eprint
  {https://arxiv.org/abs/hep-ph/0411133} {arXiv:hep-ph/0411133} \BibitemShut
  {NoStop}%
\bibitem [{\citenamefont {Cosme}\ \emph {et~al.}(2005)\citenamefont {Cosme},
  \citenamefont {Lopez~Honorez},\ and\ \citenamefont {Tytgat}}]{Cosme:2005sb}%
  \BibitemOpen
  \bibfield  {author} {\bibinfo {author} {\bibfnamefont {N.}~\bibnamefont
  {Cosme}}, \bibinfo {author} {\bibfnamefont {L.}~\bibnamefont
  {Lopez~Honorez}},\ and\ \bibinfo {author} {\bibfnamefont {M.~H.~G.}\
  \bibnamefont {Tytgat}},\ }\bibfield  {title} {\bibinfo {title} {{Leptogenesis
  and dark matter related?}},\ }\href
  {https://doi.org/10.1103/PhysRevD.72.043505} {\bibfield  {journal} {\bibinfo
  {journal} {Phys. Rev. D}\ }\textbf {\bibinfo {volume} {72}},\ \bibinfo
  {pages} {043505} (\bibinfo {year} {2005})},\ \Eprint
  {https://arxiv.org/abs/hep-ph/0506320} {arXiv:hep-ph/0506320} \BibitemShut
  {NoStop}%
\bibitem [{\citenamefont {An}\ \emph {et~al.}(2010)\citenamefont {An},
  \citenamefont {Chen}, \citenamefont {Mohapatra},\ and\ \citenamefont
  {Zhang}}]{An:2009vq}%
  \BibitemOpen
  \bibfield  {author} {\bibinfo {author} {\bibfnamefont {H.}~\bibnamefont
  {An}}, \bibinfo {author} {\bibfnamefont {S.-L.}\ \bibnamefont {Chen}},
  \bibinfo {author} {\bibfnamefont {R.~N.}\ \bibnamefont {Mohapatra}},\ and\
  \bibinfo {author} {\bibfnamefont {Y.}~\bibnamefont {Zhang}},\ }\bibfield
  {title} {\bibinfo {title} {{Leptogenesis as a Common Origin for Matter and
  Dark Matter}},\ }\href {https://doi.org/10.1007/JHEP03(2010)124} {\bibfield
  {journal} {\bibinfo  {journal} {JHEP}\ }\textbf {\bibinfo {volume} {03}},\
  \bibinfo {pages} {124}},\ \Eprint {https://arxiv.org/abs/0911.4463}
  {arXiv:0911.4463 [hep-ph]} \BibitemShut {NoStop}%
\bibitem [{\citenamefont {Hall}\ \emph
  {et~al.}(2010{\natexlab{a}})\citenamefont {Hall}, \citenamefont
  {March-Russell},\ and\ \citenamefont {West}}]{Hall:2010jx}%
  \BibitemOpen
  \bibfield  {author} {\bibinfo {author} {\bibfnamefont {L.~J.}\ \bibnamefont
  {Hall}}, \bibinfo {author} {\bibfnamefont {J.}~\bibnamefont
  {March-Russell}},\ and\ \bibinfo {author} {\bibfnamefont {S.~M.}\
  \bibnamefont {West}},\ }\bibfield  {title} {\bibinfo {title} {{A Unified
  Theory of Matter Genesis: Asymmetric Freeze-In}},\ }\href@noop {} {\
  (\bibinfo {year} {2010}{\natexlab{a}})},\ \Eprint
  {https://arxiv.org/abs/1010.0245} {arXiv:1010.0245 [hep-ph]} \BibitemShut
  {NoStop}%
\bibitem [{\citenamefont {Falkowski}\ \emph {et~al.}(2011)\citenamefont
  {Falkowski}, \citenamefont {Ruderman},\ and\ \citenamefont
  {Volansky}}]{Falkowski:2011xh}%
  \BibitemOpen
  \bibfield  {author} {\bibinfo {author} {\bibfnamefont {A.}~\bibnamefont
  {Falkowski}}, \bibinfo {author} {\bibfnamefont {J.~T.}\ \bibnamefont
  {Ruderman}},\ and\ \bibinfo {author} {\bibfnamefont {T.}~\bibnamefont
  {Volansky}},\ }\bibfield  {title} {\bibinfo {title} {{Asymmetric Dark Matter
  from Leptogenesis}},\ }\href {https://doi.org/10.1007/JHEP05(2011)106}
  {\bibfield  {journal} {\bibinfo  {journal} {JHEP}\ }\textbf {\bibinfo
  {volume} {05}},\ \bibinfo {pages} {106}},\ \Eprint
  {https://arxiv.org/abs/1101.4936} {arXiv:1101.4936 [hep-ph]} \BibitemShut
  {NoStop}%
\bibitem [{\citenamefont {Chun}(2011)}]{Chun:2011cc}%
  \BibitemOpen
  \bibfield  {author} {\bibinfo {author} {\bibfnamefont {E.~J.}\ \bibnamefont
  {Chun}},\ }\bibfield  {title} {\bibinfo {title} {{Minimal Dark Matter and
  Leptogenesis}},\ }\href {https://doi.org/10.1007/JHEP03(2011)098} {\bibfield
  {journal} {\bibinfo  {journal} {JHEP}\ }\textbf {\bibinfo {volume} {03}},\
  \bibinfo {pages} {098}},\ \Eprint {https://arxiv.org/abs/1102.3455}
  {arXiv:1102.3455 [hep-ph]} \BibitemShut {NoStop}%
\bibitem [{\citenamefont {Feng}\ \emph {et~al.}(2012)\citenamefont {Feng},
  \citenamefont {Nath},\ and\ \citenamefont {Peim}}]{Feng:2012jn}%
  \BibitemOpen
  \bibfield  {author} {\bibinfo {author} {\bibfnamefont {W.-Z.}\ \bibnamefont
  {Feng}}, \bibinfo {author} {\bibfnamefont {P.}~\bibnamefont {Nath}},\ and\
  \bibinfo {author} {\bibfnamefont {G.}~\bibnamefont {Peim}},\ }\bibfield
  {title} {\bibinfo {title} {{Cosmic Coincidence and Asymmetric Dark Matter in
  a Stueckelberg Extension}},\ }\href
  {https://doi.org/10.1103/PhysRevD.85.115016} {\bibfield  {journal} {\bibinfo
  {journal} {Phys. Rev. D}\ }\textbf {\bibinfo {volume} {85}},\ \bibinfo
  {pages} {115016} (\bibinfo {year} {2012})},\ \Eprint
  {https://arxiv.org/abs/1204.5752} {arXiv:1204.5752 [hep-ph]} \BibitemShut
  {NoStop}%
\bibitem [{\citenamefont {Unwin}(2014)}]{Unwin:2014poa}%
  \BibitemOpen
  \bibfield  {author} {\bibinfo {author} {\bibfnamefont {J.}~\bibnamefont
  {Unwin}},\ }\bibfield  {title} {\bibinfo {title} {{Towards cogenesis via
  Asymmetric Freeze-In: The $\chi$ who came-in from the cold}},\ }\href
  {https://doi.org/10.1007/JHEP10(2014)190} {\bibfield  {journal} {\bibinfo
  {journal} {JHEP}\ }\textbf {\bibinfo {volume} {10}},\ \bibinfo {pages}
  {190}},\ \Eprint {https://arxiv.org/abs/1406.3027} {arXiv:1406.3027 [hep-ph]}
  \BibitemShut {NoStop}%
\bibitem [{\citenamefont {Di~Bari}\ \emph {et~al.}(2016)\citenamefont
  {Di~Bari}, \citenamefont {Ludl},\ and\ \citenamefont
  {Palomares-Ruiz}}]{DiBari:2016guw}%
  \BibitemOpen
  \bibfield  {author} {\bibinfo {author} {\bibfnamefont {P.}~\bibnamefont
  {Di~Bari}}, \bibinfo {author} {\bibfnamefont {P.~O.}\ \bibnamefont {Ludl}},\
  and\ \bibinfo {author} {\bibfnamefont {S.}~\bibnamefont {Palomares-Ruiz}},\
  }\bibfield  {title} {\bibinfo {title} {{Unifying leptogenesis, dark matter
  and high-energy neutrinos with right-handed neutrino mixing via Higgs
  portal}},\ }\href {https://doi.org/10.1088/1475-7516/2016/11/044} {\bibfield
  {journal} {\bibinfo  {journal} {JCAP}\ }\textbf {\bibinfo {volume} {11}},\
  \bibinfo {pages} {044}},\ \Eprint {https://arxiv.org/abs/1606.06238}
  {arXiv:1606.06238 [hep-ph]} \BibitemShut {NoStop}%
\bibitem [{\citenamefont {Escudero}\ \emph
  {et~al.}(2017{\natexlab{a}})\citenamefont {Escudero}, \citenamefont {Rius},\
  and\ \citenamefont {Sanz}}]{Escudero:2016tzx}%
  \BibitemOpen
  \bibfield  {author} {\bibinfo {author} {\bibfnamefont {M.}~\bibnamefont
  {Escudero}}, \bibinfo {author} {\bibfnamefont {N.}~\bibnamefont {Rius}},\
  and\ \bibinfo {author} {\bibfnamefont {V.}~\bibnamefont {Sanz}},\ }\bibfield
  {title} {\bibinfo {title} {{Sterile neutrino portal to Dark Matter I: The
  $U(1)_{B-L}$ case}},\ }\href {https://doi.org/10.1007/JHEP02(2017)045}
  {\bibfield  {journal} {\bibinfo  {journal} {JHEP}\ }\textbf {\bibinfo
  {volume} {02}},\ \bibinfo {pages} {045}},\ \Eprint
  {https://arxiv.org/abs/1606.01258} {arXiv:1606.01258 [hep-ph]} \BibitemShut
  {NoStop}%
\bibitem [{\citenamefont {Escudero}\ \emph
  {et~al.}(2017{\natexlab{b}})\citenamefont {Escudero}, \citenamefont {Rius},\
  and\ \citenamefont {Sanz}}]{Escudero:2016ksa}%
  \BibitemOpen
  \bibfield  {author} {\bibinfo {author} {\bibfnamefont {M.}~\bibnamefont
  {Escudero}}, \bibinfo {author} {\bibfnamefont {N.}~\bibnamefont {Rius}},\
  and\ \bibinfo {author} {\bibfnamefont {V.}~\bibnamefont {Sanz}},\ }\bibfield
  {title} {\bibinfo {title} {{Sterile Neutrino portal to Dark Matter II: Exact
  Dark symmetry}},\ }\href {https://doi.org/10.1140/epjc/s10052-017-4963-x}
  {\bibfield  {journal} {\bibinfo  {journal} {Eur. Phys. J. C}\ }\textbf
  {\bibinfo {volume} {77}},\ \bibinfo {pages} {397} (\bibinfo {year}
  {2017}{\natexlab{b}})},\ \Eprint {https://arxiv.org/abs/1607.02373}
  {arXiv:1607.02373 [hep-ph]} \BibitemShut {NoStop}%
\bibitem [{\citenamefont {Gonz\'alez-Mac\'\i{}as}\ \emph
  {et~al.}(2016)\citenamefont {Gonz\'alez-Mac\'\i{}as}, \citenamefont
  {Illana},\ and\ \citenamefont {Wudka}}]{Gonzalez-Macias:2016vxy}%
  \BibitemOpen
  \bibfield  {author} {\bibinfo {author} {\bibfnamefont {V.}~\bibnamefont
  {Gonz\'alez-Mac\'\i{}as}}, \bibinfo {author} {\bibfnamefont {J.~I.}\
  \bibnamefont {Illana}},\ and\ \bibinfo {author} {\bibfnamefont
  {J.}~\bibnamefont {Wudka}},\ }\bibfield  {title} {\bibinfo {title} {{A
  realistic model for Dark Matter interactions in the neutrino portal
  paradigm}},\ }\href {https://doi.org/10.1007/JHEP05(2016)171} {\bibfield
  {journal} {\bibinfo  {journal} {JHEP}\ }\textbf {\bibinfo {volume} {05}},\
  \bibinfo {pages} {171}},\ \Eprint {https://arxiv.org/abs/1601.05051}
  {arXiv:1601.05051 [hep-ph]} \BibitemShut {NoStop}%
\bibitem [{\citenamefont {Ballesteros}\ \emph {et~al.}(2017)\citenamefont
  {Ballesteros}, \citenamefont {Redondo}, \citenamefont {Ringwald},\ and\
  \citenamefont {Tamarit}}]{Ballesteros_2017}%
  \BibitemOpen
  \bibfield  {author} {\bibinfo {author} {\bibfnamefont {G.}~\bibnamefont
  {Ballesteros}}, \bibinfo {author} {\bibfnamefont {J.}~\bibnamefont
  {Redondo}}, \bibinfo {author} {\bibfnamefont {A.}~\bibnamefont {Ringwald}},\
  and\ \bibinfo {author} {\bibfnamefont {C.}~\bibnamefont {Tamarit}},\
  }\bibfield  {title} {\bibinfo {title} {Standard
  model—axion—seesaw—higgs portal inflation. five problems of particle
  physics and cosmology solved in one stroke},\ }\href
  {https://doi.org/10.1088/1475-7516/2017/08/001} {\bibfield  {journal}
  {\bibinfo  {journal} {Journal of Cosmology and Astroparticle Physics}\
  }\textbf {\bibinfo {volume} {2017}}\bibinfo  {number} { (08)},\ \bibinfo
  {pages} {001–001}}\BibitemShut {NoStop}%
\bibitem [{\citenamefont {Falkowski}\ \emph {et~al.}(2019)\citenamefont
  {Falkowski}, \citenamefont {Kuflik}, \citenamefont {Levi},\ and\
  \citenamefont {Volansky}}]{Falkowski:2017uya}%
  \BibitemOpen
\bibfield  {number} {  }\bibfield  {author} {\bibinfo {author} {\bibfnamefont
  {A.}~\bibnamefont {Falkowski}}, \bibinfo {author} {\bibfnamefont
  {E.}~\bibnamefont {Kuflik}}, \bibinfo {author} {\bibfnamefont
  {N.}~\bibnamefont {Levi}},\ and\ \bibinfo {author} {\bibfnamefont
  {T.}~\bibnamefont {Volansky}},\ }\bibfield  {title} {\bibinfo {title} {{Light
  Dark Matter from Leptogenesis}},\ }\href
  {https://doi.org/10.1103/PhysRevD.99.015022} {\bibfield  {journal} {\bibinfo
  {journal} {Phys. Rev. D}\ }\textbf {\bibinfo {volume} {99}},\ \bibinfo
  {pages} {015022} (\bibinfo {year} {2019})},\ \Eprint
  {https://arxiv.org/abs/1712.07652} {arXiv:1712.07652 [hep-ph]} \BibitemShut
  {NoStop}%
\bibitem [{\citenamefont {Biswas}\ \emph {et~al.}(2019)\citenamefont {Biswas},
  \citenamefont {Choubey}, \citenamefont {Covi},\ and\ \citenamefont
  {Khan}}]{Biswas:2018sib}%
  \BibitemOpen
  \bibfield  {author} {\bibinfo {author} {\bibfnamefont {A.}~\bibnamefont
  {Biswas}}, \bibinfo {author} {\bibfnamefont {S.}~\bibnamefont {Choubey}},
  \bibinfo {author} {\bibfnamefont {L.}~\bibnamefont {Covi}},\ and\ \bibinfo
  {author} {\bibfnamefont {S.}~\bibnamefont {Khan}},\ }\bibfield  {title}
  {\bibinfo {title} {{Common origin of baryon asymmetry, dark matter and
  neutrino mass}},\ }\href {https://doi.org/10.1007/JHEP05(2019)193} {\bibfield
   {journal} {\bibinfo  {journal} {JHEP}\ }\textbf {\bibinfo {volume} {05}},\
  \bibinfo {pages} {193}},\ \Eprint {https://arxiv.org/abs/1812.06122}
  {arXiv:1812.06122 [hep-ph]} \BibitemShut {NoStop}%
\bibitem [{\citenamefont {Caputo}\ \emph {et~al.}(2019)\citenamefont {Caputo},
  \citenamefont {Hernandez},\ and\ \citenamefont {Rius}}]{Caputo:2018zky}%
  \BibitemOpen
  \bibfield  {author} {\bibinfo {author} {\bibfnamefont {A.}~\bibnamefont
  {Caputo}}, \bibinfo {author} {\bibfnamefont {P.}~\bibnamefont {Hernandez}},\
  and\ \bibinfo {author} {\bibfnamefont {N.}~\bibnamefont {Rius}},\ }\bibfield
  {title} {\bibinfo {title} {{Leptogenesis from oscillations and dark
  matter}},\ }\href {https://doi.org/10.1140/epjc/s10052-019-7083-y} {\bibfield
   {journal} {\bibinfo  {journal} {Eur. Phys. J. C}\ }\textbf {\bibinfo
  {volume} {79}},\ \bibinfo {pages} {574} (\bibinfo {year} {2019})},\ \Eprint
  {https://arxiv.org/abs/1807.03309} {arXiv:1807.03309 [hep-ph]} \BibitemShut
  {NoStop}%
\bibitem [{\citenamefont {Chianese}\ \emph {et~al.}(2020)\citenamefont
  {Chianese}, \citenamefont {Fu},\ and\ \citenamefont
  {King}}]{Chianese:2019epo}%
  \BibitemOpen
  \bibfield  {author} {\bibinfo {author} {\bibfnamefont {M.}~\bibnamefont
  {Chianese}}, \bibinfo {author} {\bibfnamefont {B.}~\bibnamefont {Fu}},\ and\
  \bibinfo {author} {\bibfnamefont {S.~F.}\ \bibnamefont {King}},\ }\bibfield
  {title} {\bibinfo {title} {{Minimal Seesaw extension for Neutrino Mass and
  Mixing, Leptogenesis and Dark Matter: FIMPzillas through the Right-Handed
  Neutrino Portal}},\ }\href {https://doi.org/10.1088/1475-7516/2020/03/030}
  {\bibfield  {journal} {\bibinfo  {journal} {JCAP}\ }\textbf {\bibinfo
  {volume} {03}},\ \bibinfo {pages} {030}},\ \Eprint
  {https://arxiv.org/abs/1910.12916} {arXiv:1910.12916 [hep-ph]} \BibitemShut
  {NoStop}%
\bibitem [{\citenamefont {Bandyopadhyay}\ \emph {et~al.}(2020)\citenamefont
  {Bandyopadhyay}, \citenamefont {Chun},\ and\ \citenamefont
  {Mandal}}]{Bandyopadhyay_2020}%
  \BibitemOpen
  \bibfield  {author} {\bibinfo {author} {\bibfnamefont {P.}~\bibnamefont
  {Bandyopadhyay}}, \bibinfo {author} {\bibfnamefont {E.~J.}\ \bibnamefont
  {Chun}},\ and\ \bibinfo {author} {\bibfnamefont {R.}~\bibnamefont {Mandal}},\
  }\bibfield  {title} {\bibinfo {title} {Feeble neutrino portal dark matter at
  neutrino detectors},\ }\href {https://doi.org/10.1088/1475-7516/2020/08/019}
  {\bibfield  {journal} {\bibinfo  {journal} {Journal of Cosmology and
  Astroparticle Physics}\ }\textbf {\bibinfo {volume} {2020}}\bibinfo  {number}
  { (08)},\ \bibinfo {pages} {019–019}}\BibitemShut {NoStop}%
\bibitem [{\citenamefont {Coy}\ \emph {et~al.}(2021)\citenamefont {Coy},
  \citenamefont {Gupta},\ and\ \citenamefont {Hambye}}]{Coy_2021}%
  \BibitemOpen
\bibfield  {number} {  }\bibfield  {author} {\bibinfo {author} {\bibfnamefont
  {R.}~\bibnamefont {Coy}}, \bibinfo {author} {\bibfnamefont {A.}~\bibnamefont
  {Gupta}},\ and\ \bibinfo {author} {\bibfnamefont {T.}~\bibnamefont
  {Hambye}},\ }\bibfield  {title} {\bibinfo {title} {Seesaw neutrino
  determination of the dark matter relic density},\ }\bibfield  {journal}
  {\bibinfo  {journal} {Physical Review D}\ }\textbf {\bibinfo {volume}
  {104}},\ \href {https://doi.org/10.1103/physrevd.104.083024}
  {10.1103/physrevd.104.083024} (\bibinfo {year} {2021})\BibitemShut {NoStop}%
\bibitem [{\citenamefont {Datta}\ \emph {et~al.}(2021)\citenamefont {Datta},
  \citenamefont {Roshan},\ and\ \citenamefont {Sil}}]{Datta:2021elq}%
  \BibitemOpen
  \bibfield  {author} {\bibinfo {author} {\bibfnamefont {A.}~\bibnamefont
  {Datta}}, \bibinfo {author} {\bibfnamefont {R.}~\bibnamefont {Roshan}},\ and\
  \bibinfo {author} {\bibfnamefont {A.}~\bibnamefont {Sil}},\ }\bibfield
  {title} {\bibinfo {title} {{Imprint of the Seesaw Mechanism on Feebly
  Interacting Dark Matter and the Baryon Asymmetry}},\ }\href
  {https://doi.org/10.1103/PhysRevLett.127.231801} {\bibfield  {journal}
  {\bibinfo  {journal} {Phys. Rev. Lett.}\ }\textbf {\bibinfo {volume} {127}},\
  \bibinfo {pages} {231801} (\bibinfo {year} {2021})},\ \Eprint
  {https://arxiv.org/abs/2104.02030} {arXiv:2104.02030 [hep-ph]} \BibitemShut
  {NoStop}%
\bibitem [{\citenamefont {Bhattacharya}\ \emph {et~al.}(2022)\citenamefont
  {Bhattacharya}, \citenamefont {Roshan}, \citenamefont {Sil},\ and\
  \citenamefont {Vatsyayan}}]{Bhattacharya:2021jli}%
  \BibitemOpen
  \bibfield  {author} {\bibinfo {author} {\bibfnamefont {S.}~\bibnamefont
  {Bhattacharya}}, \bibinfo {author} {\bibfnamefont {R.}~\bibnamefont
  {Roshan}}, \bibinfo {author} {\bibfnamefont {A.}~\bibnamefont {Sil}},\ and\
  \bibinfo {author} {\bibfnamefont {D.}~\bibnamefont {Vatsyayan}},\ }\bibfield
  {title} {\bibinfo {title} {{Symmetry origin of baryon asymmetry, dark matter,
  and neutrino mass}},\ }\href {https://doi.org/10.1103/PhysRevD.106.075005}
  {\bibfield  {journal} {\bibinfo  {journal} {Phys. Rev. D}\ }\textbf {\bibinfo
  {volume} {106}},\ \bibinfo {pages} {075005} (\bibinfo {year} {2022})},\
  \Eprint {https://arxiv.org/abs/2105.06189} {arXiv:2105.06189 [hep-ph]}
  \BibitemShut {NoStop}%
\bibitem [{\citenamefont {Sopov}\ and\ \citenamefont
  {Volkas}(2023)}]{Sopov:2022bog}%
  \BibitemOpen
  \bibfield  {author} {\bibinfo {author} {\bibfnamefont {A.~H.}\ \bibnamefont
  {Sopov}}\ and\ \bibinfo {author} {\bibfnamefont {R.~R.}\ \bibnamefont
  {Volkas}},\ }\bibfield  {title} {\bibinfo {title} {{VISH\ensuremath{\nu}:
  solving five Standard Model shortcomings with a Poincar\'e-protected
  electroweak scale}},\ }\href {https://doi.org/10.1016/j.dark.2023.101381}
  {\bibfield  {journal} {\bibinfo  {journal} {Phys. Dark Univ.}\ }\textbf
  {\bibinfo {volume} {42}},\ \bibinfo {pages} {101381} (\bibinfo {year}
  {2023})},\ \Eprint {https://arxiv.org/abs/2206.11598} {arXiv:2206.11598
  [hep-ph]} \BibitemShut {NoStop}%
\bibitem [{\citenamefont {Coito}\ \emph {et~al.}(2022)\citenamefont {Coito},
  \citenamefont {Faubel}, \citenamefont {Herrero-Garc\'\i{}a}, \citenamefont
  {Santamaria},\ and\ \citenamefont {Titov}}]{Coito:2022kif}%
  \BibitemOpen
  \bibfield  {author} {\bibinfo {author} {\bibfnamefont {L.}~\bibnamefont
  {Coito}}, \bibinfo {author} {\bibfnamefont {C.}~\bibnamefont {Faubel}},
  \bibinfo {author} {\bibfnamefont {J.}~\bibnamefont {Herrero-Garc\'\i{}a}},
  \bibinfo {author} {\bibfnamefont {A.}~\bibnamefont {Santamaria}},\ and\
  \bibinfo {author} {\bibfnamefont {A.}~\bibnamefont {Titov}},\ }\bibfield
  {title} {\bibinfo {title} {{Sterile neutrino portals to Majorana dark matter:
  effective operators and UV completions}},\ }\href
  {https://doi.org/10.1007/JHEP08(2022)085} {\bibfield  {journal} {\bibinfo
  {journal} {JHEP}\ }\textbf {\bibinfo {volume} {08}},\ \bibinfo {pages}
  {085}},\ \Eprint {https://arxiv.org/abs/2203.01946} {arXiv:2203.01946
  [hep-ph]} \BibitemShut {NoStop}%
\bibitem [{\citenamefont {Berbig}\ and\ \citenamefont
  {Ghoshal}(2023)}]{Berbig:2023yyy}%
  \BibitemOpen
  \bibfield  {author} {\bibinfo {author} {\bibfnamefont {M.}~\bibnamefont
  {Berbig}}\ and\ \bibinfo {author} {\bibfnamefont {A.}~\bibnamefont
  {Ghoshal}},\ }\bibfield  {title} {\bibinfo {title} {{Impact of high-scale
  Seesaw and Leptogenesis on inflationary tensor perturbations as detectable
  gravitational waves}},\ }\href {https://doi.org/10.1007/JHEP05(2023)172}
  {\bibfield  {journal} {\bibinfo  {journal} {JHEP}\ }\textbf {\bibinfo
  {volume} {05}},\ \bibinfo {pages} {172}},\ \Eprint
  {https://arxiv.org/abs/2301.05672} {arXiv:2301.05672 [hep-ph]} \BibitemShut
  {NoStop}%
\bibitem [{\citenamefont {Li}\ and\ \citenamefont {Xu}(2023)}]{Li_2023}%
  \BibitemOpen
  \bibfield  {author} {\bibinfo {author} {\bibfnamefont {S.-P.}\ \bibnamefont
  {Li}}\ and\ \bibinfo {author} {\bibfnamefont {X.-J.}\ \bibnamefont {Xu}},\
  }\bibfield  {title} {\bibinfo {title} {Dark matter produced from right-handed
  neutrinos},\ }\href {https://doi.org/10.1088/1475-7516/2023/06/047}
  {\bibfield  {journal} {\bibinfo  {journal} {Journal of Cosmology and
  Astroparticle Physics}\ }\textbf {\bibinfo {volume} {2023}}\bibinfo  {number}
  { (06)},\ \bibinfo {pages} {047}}\BibitemShut {NoStop}%
\bibitem [{\citenamefont {Barman}\ \emph {et~al.}(2023)\citenamefont {Barman},
  \citenamefont {Dev},\ and\ \citenamefont
  {Ghoshal}}]{barman2023probingfreezeindarkmatter}%
  \BibitemOpen
\bibfield  {number} {  }\bibfield  {author} {\bibinfo {author} {\bibfnamefont
  {B.}~\bibnamefont {Barman}}, \bibinfo {author} {\bibfnamefont {P.~S.~B.}\
  \bibnamefont {Dev}},\ and\ \bibinfo {author} {\bibfnamefont {A.}~\bibnamefont
  {Ghoshal}},\ }\href {https://arxiv.org/abs/2210.07739} {\bibinfo {title}
  {Probing freeze-in dark matter via heavy neutrino portal}} (\bibinfo {year}
  {2023}),\ \Eprint {https://arxiv.org/abs/2210.07739} {arXiv:2210.07739
  [hep-ph]} \BibitemShut {NoStop}%
\bibitem [{\citenamefont {Herrero-Garcia}\ \emph {et~al.}(2023)\citenamefont
  {Herrero-Garcia}, \citenamefont {Landini},\ and\ \citenamefont
  {Vatsyayan}}]{Herrero-Garcia:2023lhv}%
  \BibitemOpen
  \bibfield  {author} {\bibinfo {author} {\bibfnamefont {J.}~\bibnamefont
  {Herrero-Garcia}}, \bibinfo {author} {\bibfnamefont {G.}~\bibnamefont
  {Landini}},\ and\ \bibinfo {author} {\bibfnamefont {D.}~\bibnamefont
  {Vatsyayan}},\ }\bibfield  {title} {\bibinfo {title} {{Asymmetries in
  extended dark sectors: a cogenesis scenario}},\ }\href
  {https://doi.org/10.1007/JHEP05(2023)049} {\bibfield  {journal} {\bibinfo
  {journal} {JHEP}\ }\textbf {\bibinfo {volume} {05}},\ \bibinfo {pages}
  {049}},\ \Eprint {https://arxiv.org/abs/2301.13238} {arXiv:2301.13238
  [hep-ph]} \BibitemShut {NoStop}%
\bibitem [{\citenamefont {Berbig}\ \emph {et~al.}(2024)\citenamefont {Berbig},
  \citenamefont {Herrero-Garcia},\ and\ \citenamefont
  {Landini}}]{Berbig:2024uwm}%
  \BibitemOpen
  \bibfield  {author} {\bibinfo {author} {\bibfnamefont {M.}~\bibnamefont
  {Berbig}}, \bibinfo {author} {\bibfnamefont {J.}~\bibnamefont
  {Herrero-Garcia}},\ and\ \bibinfo {author} {\bibfnamefont {G.}~\bibnamefont
  {Landini}},\ }\bibfield  {title} {\bibinfo {title} {{Dynamical origin of
  neutrino masses and dark matter from a new confining sector}},\ }\href
  {https://doi.org/10.1103/PhysRevD.110.035011} {\bibfield  {journal} {\bibinfo
   {journal} {Phys. Rev. D}\ }\textbf {\bibinfo {volume} {110}},\ \bibinfo
  {pages} {035011} (\bibinfo {year} {2024})},\ \Eprint
  {https://arxiv.org/abs/2403.17488} {arXiv:2403.17488 [hep-ph]} \BibitemShut
  {NoStop}%
\bibitem [{\citenamefont {Cline}\ \emph {et~al.}(2013)\citenamefont {Cline},
  \citenamefont {Scott}, \citenamefont {Kainulainen},\ and\ \citenamefont
  {Weniger}}]{Cline_2013}%
  \BibitemOpen
  \bibfield  {author} {\bibinfo {author} {\bibfnamefont {J.~M.}\ \bibnamefont
  {Cline}}, \bibinfo {author} {\bibfnamefont {P.}~\bibnamefont {Scott}},
  \bibinfo {author} {\bibfnamefont {K.}~\bibnamefont {Kainulainen}},\ and\
  \bibinfo {author} {\bibfnamefont {C.}~\bibnamefont {Weniger}},\ }\bibfield
  {title} {\bibinfo {title} {Update on scalar singlet dark matter},\ }\bibfield
   {journal} {\bibinfo  {journal} {Physical Review D}\ }\textbf {\bibinfo
  {volume} {88}},\ \href {https://doi.org/10.1103/physrevd.88.055025}
  {10.1103/physrevd.88.055025} (\bibinfo {year} {2013})\BibitemShut {NoStop}%
\bibitem [{\citenamefont {Athron}\ \emph {et~al.}(2017)\citenamefont {Athron}
  \emph {et~al.}}]{GAMBIT:2017gge}%
  \BibitemOpen
  \bibfield  {author} {\bibinfo {author} {\bibfnamefont {P.}~\bibnamefont
  {Athron}} \emph {et~al.} (\bibinfo {collaboration} {GAMBIT}),\ }\bibfield
  {title} {\bibinfo {title} {{Status of the scalar singlet dark matter
  model}},\ }\href {https://doi.org/10.1140/epjc/s10052-017-5113-1} {\bibfield
  {journal} {\bibinfo  {journal} {Eur. Phys. J. C}\ }\textbf {\bibinfo {volume}
  {77}},\ \bibinfo {pages} {568} (\bibinfo {year} {2017})},\ \Eprint
  {https://arxiv.org/abs/1705.07931} {arXiv:1705.07931 [hep-ph]} \BibitemShut
  {NoStop}%
\bibitem [{\citenamefont {Griest}\ and\ \citenamefont
  {Kamionkowski}(1990)}]{Griest:1989wd}%
  \BibitemOpen
  \bibfield  {author} {\bibinfo {author} {\bibfnamefont {K.}~\bibnamefont
  {Griest}}\ and\ \bibinfo {author} {\bibfnamefont {M.}~\bibnamefont
  {Kamionkowski}},\ }\bibfield  {title} {\bibinfo {title} {{Unitarity Limits on
  the Mass and Radius of Dark Matter Particles}},\ }\href
  {https://doi.org/10.1103/PhysRevLett.64.615} {\bibfield  {journal} {\bibinfo
  {journal} {Phys. Rev. Lett.}\ }\textbf {\bibinfo {volume} {64}},\ \bibinfo
  {pages} {615} (\bibinfo {year} {1990})}\BibitemShut {NoStop}%
\bibitem [{\citenamefont {Hall}\ \emph
  {et~al.}(2010{\natexlab{b}})\citenamefont {Hall}, \citenamefont {Jedamzik},
  \citenamefont {March-Russell},\ and\ \citenamefont {West}}]{Hall:2009bx}%
  \BibitemOpen
  \bibfield  {author} {\bibinfo {author} {\bibfnamefont {L.~J.}\ \bibnamefont
  {Hall}}, \bibinfo {author} {\bibfnamefont {K.}~\bibnamefont {Jedamzik}},
  \bibinfo {author} {\bibfnamefont {J.}~\bibnamefont {March-Russell}},\ and\
  \bibinfo {author} {\bibfnamefont {S.~M.}\ \bibnamefont {West}},\ }\bibfield
  {title} {\bibinfo {title} {{Freeze-In Production of FIMP Dark Matter}},\
  }\href {https://doi.org/10.1007/JHEP03(2010)080} {\bibfield  {journal}
  {\bibinfo  {journal} {JHEP}\ }\textbf {\bibinfo {volume} {03}},\ \bibinfo
  {pages} {080}},\ \Eprint {https://arxiv.org/abs/0911.1120} {arXiv:0911.1120
  [hep-ph]} \BibitemShut {NoStop}%
\bibitem [{\citenamefont {Bernal}\ \emph {et~al.}(2017)\citenamefont {Bernal},
  \citenamefont {Heikinheimo}, \citenamefont {Tenkanen}, \citenamefont
  {Tuominen},\ and\ \citenamefont {Vaskonen}}]{Bernal_2017}%
  \BibitemOpen
  \bibfield  {author} {\bibinfo {author} {\bibfnamefont {N.}~\bibnamefont
  {Bernal}}, \bibinfo {author} {\bibfnamefont {M.}~\bibnamefont {Heikinheimo}},
  \bibinfo {author} {\bibfnamefont {T.}~\bibnamefont {Tenkanen}}, \bibinfo
  {author} {\bibfnamefont {K.}~\bibnamefont {Tuominen}},\ and\ \bibinfo
  {author} {\bibfnamefont {V.}~\bibnamefont {Vaskonen}},\ }\bibfield  {title}
  {\bibinfo {title} {The dawn of fimp dark matter: A review of models and
  constraints},\ }\href {https://doi.org/10.1142/s0217751x1730023x} {\bibfield
  {journal} {\bibinfo  {journal} {International Journal of Modern Physics A}\
  }\textbf {\bibinfo {volume} {32}},\ \bibinfo {pages} {1730023} (\bibinfo
  {year} {2017})}\BibitemShut {NoStop}%
\bibitem [{\citenamefont {Vattis}\ \emph {et~al.}(2019)\citenamefont {Vattis},
  \citenamefont {Koushiappas},\ and\ \citenamefont {Loeb}}]{Vattis_2019}%
  \BibitemOpen
  \bibfield  {author} {\bibinfo {author} {\bibfnamefont {K.}~\bibnamefont
  {Vattis}}, \bibinfo {author} {\bibfnamefont {S.~M.}\ \bibnamefont
  {Koushiappas}},\ and\ \bibinfo {author} {\bibfnamefont {A.}~\bibnamefont
  {Loeb}},\ }\bibfield  {title} {\bibinfo {title} {Dark matter decaying in the
  late universe can relieve the $h_0$ tension},\ }\bibfield  {journal}
  {\bibinfo  {journal} {Physical Review D}\ }\textbf {\bibinfo {volume} {99}},\
  \href {https://doi.org/10.1103/physrevd.99.121302}
  {10.1103/physrevd.99.121302} (\bibinfo {year} {2019})\BibitemShut {NoStop}%
\bibitem [{\citenamefont {Di~Valentino}\ \emph {et~al.}(2021)\citenamefont
  {Di~Valentino}, \citenamefont {Mena}, \citenamefont {Pan}, \citenamefont
  {Visinelli}, \citenamefont {Yang}, \citenamefont {Melchiorri}, \citenamefont
  {Mota}, \citenamefont {Riess},\ and\ \citenamefont
  {Silk}}]{DiValentino:2021izs}%
  \BibitemOpen
  \bibfield  {author} {\bibinfo {author} {\bibfnamefont {E.}~\bibnamefont
  {Di~Valentino}}, \bibinfo {author} {\bibfnamefont {O.}~\bibnamefont {Mena}},
  \bibinfo {author} {\bibfnamefont {S.}~\bibnamefont {Pan}}, \bibinfo {author}
  {\bibfnamefont {L.}~\bibnamefont {Visinelli}}, \bibinfo {author}
  {\bibfnamefont {W.}~\bibnamefont {Yang}}, \bibinfo {author} {\bibfnamefont
  {A.}~\bibnamefont {Melchiorri}}, \bibinfo {author} {\bibfnamefont {D.~F.}\
  \bibnamefont {Mota}}, \bibinfo {author} {\bibfnamefont {A.~G.}\ \bibnamefont
  {Riess}},\ and\ \bibinfo {author} {\bibfnamefont {J.}~\bibnamefont {Silk}},\
  }\bibfield  {title} {\bibinfo {title} {{In the realm of the Hubble
  tension\textemdash{}a review of solutions}},\ }\href
  {https://doi.org/10.1088/1361-6382/ac086d} {\bibfield  {journal} {\bibinfo
  {journal} {Class. Quant. Grav.}\ }\textbf {\bibinfo {volume} {38}},\ \bibinfo
  {pages} {153001} (\bibinfo {year} {2021})},\ \Eprint
  {https://arxiv.org/abs/2103.01183} {arXiv:2103.01183 [astro-ph.CO]}
  \BibitemShut {NoStop}%
\bibitem [{\citenamefont {Schöneberg}\ \emph {et~al.}(2022)\citenamefont
  {Schöneberg}, \citenamefont {Abellán}, \citenamefont {Sánchez},
  \citenamefont {Witte}, \citenamefont {Poulin},\ and\ \citenamefont
  {Lesgourgues}}]{Sch_neberg_2022}%
  \BibitemOpen
  \bibfield  {author} {\bibinfo {author} {\bibfnamefont {N.}~\bibnamefont
  {Schöneberg}}, \bibinfo {author} {\bibfnamefont {G.~F.}\ \bibnamefont
  {Abellán}}, \bibinfo {author} {\bibfnamefont {A.~P.}\ \bibnamefont
  {Sánchez}}, \bibinfo {author} {\bibfnamefont {S.~J.}\ \bibnamefont {Witte}},
  \bibinfo {author} {\bibfnamefont {V.}~\bibnamefont {Poulin}},\ and\ \bibinfo
  {author} {\bibfnamefont {J.}~\bibnamefont {Lesgourgues}},\ }\bibfield
  {title} {\bibinfo {title} {The h0 olympics: A fair ranking of proposed
  models},\ }\href {https://doi.org/10.1016/j.physrep.2022.07.001} {\bibfield
  {journal} {\bibinfo  {journal} {Physics Reports}\ }\textbf {\bibinfo {volume}
  {984}},\ \bibinfo {pages} {1–55} (\bibinfo {year} {2022})}\BibitemShut
  {NoStop}%
\bibitem [{\citenamefont {Feng}\ \emph {et~al.}(2003)\citenamefont {Feng},
  \citenamefont {Rajaraman},\ and\ \citenamefont {Takayama}}]{Feng_2003}%
  \BibitemOpen
  \bibfield  {author} {\bibinfo {author} {\bibfnamefont {J.~L.}\ \bibnamefont
  {Feng}}, \bibinfo {author} {\bibfnamefont {A.}~\bibnamefont {Rajaraman}},\
  and\ \bibinfo {author} {\bibfnamefont {F.}~\bibnamefont {Takayama}},\
  }\bibfield  {title} {\bibinfo {title} {Superweakly interacting massive
  particles},\ }\bibfield  {journal} {\bibinfo  {journal} {Physical Review
  Letters}\ }\textbf {\bibinfo {volume} {91}},\ \href
  {https://doi.org/10.1103/physrevlett.91.011302}
  {10.1103/physrevlett.91.011302} (\bibinfo {year} {2003})\BibitemShut
  {NoStop}%
\bibitem [{\citenamefont {Feng}\ \emph {et~al.}(2004)\citenamefont {Feng},
  \citenamefont {Su},\ and\ \citenamefont {Takayama}}]{Feng_2004}%
  \BibitemOpen
  \bibfield  {author} {\bibinfo {author} {\bibfnamefont {J.~L.}\ \bibnamefont
  {Feng}}, \bibinfo {author} {\bibfnamefont {S.}~\bibnamefont {Su}},\ and\
  \bibinfo {author} {\bibfnamefont {F.}~\bibnamefont {Takayama}},\ }\bibfield
  {title} {\bibinfo {title} {Superwimp gravitino dark matter from slepton and
  sneutrino decays},\ }\bibfield  {journal} {\bibinfo  {journal} {Physical
  Review D}\ }\textbf {\bibinfo {volume} {70}},\ \href
  {https://doi.org/10.1103/physrevd.70.063514} {10.1103/physrevd.70.063514}
  (\bibinfo {year} {2004})\BibitemShut {NoStop}%
\bibitem [{\citenamefont {Cheung}\ \emph {et~al.}(2011)\citenamefont {Cheung},
  \citenamefont {Elor}, \citenamefont {Hall},\ and\ \citenamefont
  {Kumar}}]{Cheung:2010gj}%
  \BibitemOpen
  \bibfield  {author} {\bibinfo {author} {\bibfnamefont {C.}~\bibnamefont
  {Cheung}}, \bibinfo {author} {\bibfnamefont {G.}~\bibnamefont {Elor}},
  \bibinfo {author} {\bibfnamefont {L.~J.}\ \bibnamefont {Hall}},\ and\
  \bibinfo {author} {\bibfnamefont {P.}~\bibnamefont {Kumar}},\ }\bibfield
  {title} {\bibinfo {title} {{Origins of Hidden Sector Dark Matter I:
  Cosmology}},\ }\href {https://doi.org/10.1007/JHEP03(2011)042} {\bibfield
  {journal} {\bibinfo  {journal} {JHEP}\ }\textbf {\bibinfo {volume} {03}},\
  \bibinfo {pages} {042}},\ \Eprint {https://arxiv.org/abs/1010.0022}
  {arXiv:1010.0022 [hep-ph]} \BibitemShut {NoStop}%
\bibitem [{\citenamefont {Garny}\ and\ \citenamefont
  {Heisig}(2018)}]{Garny:2018ali}%
  \BibitemOpen
  \bibfield  {author} {\bibinfo {author} {\bibfnamefont {M.}~\bibnamefont
  {Garny}}\ and\ \bibinfo {author} {\bibfnamefont {J.}~\bibnamefont {Heisig}},\
  }\bibfield  {title} {\bibinfo {title} {{Interplay of super-WIMP and freeze-in
  production of dark matter}},\ }\href
  {https://doi.org/10.1103/PhysRevD.98.095031} {\bibfield  {journal} {\bibinfo
  {journal} {Phys. Rev. D}\ }\textbf {\bibinfo {volume} {98}},\ \bibinfo
  {pages} {095031} (\bibinfo {year} {2018})},\ \Eprint
  {https://arxiv.org/abs/1809.10135} {arXiv:1809.10135 [hep-ph]} \BibitemShut
  {NoStop}%
\bibitem [{\citenamefont {Aghanim}\ \emph {et~al.}(2020)\citenamefont {Aghanim}
  \emph {et~al.}}]{Planck:2018vyg}%
  \BibitemOpen
  \bibfield  {author} {\bibinfo {author} {\bibfnamefont {N.}~\bibnamefont
  {Aghanim}} \emph {et~al.} (\bibinfo {collaboration} {Planck}),\ }\bibfield
  {title} {\bibinfo {title} {{Planck 2018 results. VI. Cosmological
  parameters}},\ }\href {https://doi.org/10.1051/0004-6361/201833910}
  {\bibfield  {journal} {\bibinfo  {journal} {Astron. Astrophys.}\ }\textbf
  {\bibinfo {volume} {641}},\ \bibinfo {pages} {A6} (\bibinfo {year} {2020})},\
  \bibinfo {note} {[Erratum: Astron.Astrophys. 652, C4 (2021)]},\ \Eprint
  {https://arxiv.org/abs/1807.06209} {arXiv:1807.06209 [astro-ph.CO]}
  \BibitemShut {NoStop}%
\bibitem [{\citenamefont {Tan}\ \emph {et~al.}(2024)\citenamefont {Tan},
  \citenamefont {Dekker},\ and\ \citenamefont {Drlica-Wagner}}]{Tan:2024cek}%
  \BibitemOpen
  \bibfield  {author} {\bibinfo {author} {\bibfnamefont {C.~Y.}\ \bibnamefont
  {Tan}}, \bibinfo {author} {\bibfnamefont {A.}~\bibnamefont {Dekker}},\ and\
  \bibinfo {author} {\bibfnamefont {A.}~\bibnamefont {Drlica-Wagner}},\
  }\bibfield  {title} {\bibinfo {title} {{Mixed Warm Dark Matter Constraints
  using Milky Way Satellite Galaxy Counts}},\ }\href@noop {} {\  (\bibinfo
  {year} {2024})},\ \Eprint {https://arxiv.org/abs/2409.18917}
  {arXiv:2409.18917 [astro-ph.CO]} \BibitemShut {NoStop}%
\bibitem [{\citenamefont {Cyburt}\ \emph {et~al.}(2016)\citenamefont {Cyburt},
  \citenamefont {Fields}, \citenamefont {Olive},\ and\ \citenamefont
  {Yeh}}]{Cyburt:2015mya}%
  \BibitemOpen
  \bibfield  {author} {\bibinfo {author} {\bibfnamefont {R.~H.}\ \bibnamefont
  {Cyburt}}, \bibinfo {author} {\bibfnamefont {B.~D.}\ \bibnamefont {Fields}},
  \bibinfo {author} {\bibfnamefont {K.~A.}\ \bibnamefont {Olive}},\ and\
  \bibinfo {author} {\bibfnamefont {T.-H.}\ \bibnamefont {Yeh}},\ }\bibfield
  {title} {\bibinfo {title} {{Big Bang Nucleosynthesis: 2015}},\ }\href
  {https://doi.org/10.1103/RevModPhys.88.015004} {\bibfield  {journal}
  {\bibinfo  {journal} {Rev. Mod. Phys.}\ }\textbf {\bibinfo {volume} {88}},\
  \bibinfo {pages} {015004} (\bibinfo {year} {2016})},\ \Eprint
  {https://arxiv.org/abs/1505.01076} {arXiv:1505.01076 [astro-ph.CO]}
  \BibitemShut {NoStop}%
\bibitem [{\citenamefont {Hambye}\ \emph {et~al.}(2022)\citenamefont {Hambye},
  \citenamefont {Hufnagel},\ and\ \citenamefont {Lucca}}]{Hambye_2022}%
  \BibitemOpen
  \bibfield  {author} {\bibinfo {author} {\bibfnamefont {T.}~\bibnamefont
  {Hambye}}, \bibinfo {author} {\bibfnamefont {M.}~\bibnamefont {Hufnagel}},\
  and\ \bibinfo {author} {\bibfnamefont {M.}~\bibnamefont {Lucca}},\ }\bibfield
   {title} {\bibinfo {title} {Cosmological constraints on the decay of heavy
  relics into neutrinos},\ }\href
  {https://doi.org/10.1088/1475-7516/2022/05/033} {\bibfield  {journal}
  {\bibinfo  {journal} {Journal of Cosmology and Astroparticle Physics}\
  }\textbf {\bibinfo {volume} {2022}}\bibinfo  {number} { (05)},\ \bibinfo
  {pages} {033}}\BibitemShut {NoStop}%
\bibitem [{\citenamefont {Acharya}\ and\ \citenamefont
  {Khatri}(2020)}]{Acharya:2020gfh}%
  \BibitemOpen
\bibfield  {number} {  }\bibfield  {author} {\bibinfo {author} {\bibfnamefont
  {S.~K.}\ \bibnamefont {Acharya}}\ and\ \bibinfo {author} {\bibfnamefont
  {R.}~\bibnamefont {Khatri}},\ }\bibfield  {title} {\bibinfo {title}
  {{Constraints on $N_{\rm{eff}}$ of high energy non-thermal neutrino
  injections upto $z\sim 10^8$ from CMB spectral distortions and abundance of
  light elements}},\ }\href {https://doi.org/10.1088/1475-7516/2020/11/011}
  {\bibfield  {journal} {\bibinfo  {journal} {JCAP}\ }\textbf {\bibinfo
  {volume} {11}},\ \bibinfo {pages} {011}},\ \Eprint
  {https://arxiv.org/abs/2007.06596} {arXiv:2007.06596 [astro-ph.CO]}
  \BibitemShut {NoStop}%
\bibitem [{\citenamefont {Dror}\ \emph {et~al.}(2020)\citenamefont {Dror},
  \citenamefont {Hiramatsu}, \citenamefont {Kohri}, \citenamefont {Murayama},\
  and\ \citenamefont {White}}]{Dror_2020}%
  \BibitemOpen
  \bibfield  {author} {\bibinfo {author} {\bibfnamefont {J.~A.}\ \bibnamefont
  {Dror}}, \bibinfo {author} {\bibfnamefont {T.}~\bibnamefont {Hiramatsu}},
  \bibinfo {author} {\bibfnamefont {K.}~\bibnamefont {Kohri}}, \bibinfo
  {author} {\bibfnamefont {H.}~\bibnamefont {Murayama}},\ and\ \bibinfo
  {author} {\bibfnamefont {G.}~\bibnamefont {White}},\ }\bibfield  {title}
  {\bibinfo {title} {Testing the seesaw mechanism and leptogenesis with
  gravitational waves},\ }\bibfield  {journal} {\bibinfo  {journal} {Physical
  Review Letters}\ }\textbf {\bibinfo {volume} {124}},\ \href
  {https://doi.org/10.1103/physrevlett.124.041804}
  {10.1103/physrevlett.124.041804} (\bibinfo {year} {2020})\BibitemShut
  {NoStop}%
\bibitem [{\citenamefont {Becker}(2019)}]{Becker:2018rve}%
  \BibitemOpen
  \bibfield  {author} {\bibinfo {author} {\bibfnamefont {M.}~\bibnamefont
  {Becker}},\ }\bibfield  {title} {\bibinfo {title} {{Dark Matter from
  Freeze-In via the Neutrino Portal}},\ }\href
  {https://doi.org/10.1140/epjc/s10052-019-7095-7} {\bibfield  {journal}
  {\bibinfo  {journal} {Eur. Phys. J. C}\ }\textbf {\bibinfo {volume} {79}},\
  \bibinfo {pages} {611} (\bibinfo {year} {2019})},\ \Eprint
  {https://arxiv.org/abs/1806.08579} {arXiv:1806.08579 [hep-ph]} \BibitemShut
  {NoStop}%
\bibitem [{\citenamefont {Martin}(1992)}]{Martin:1992mq}%
  \BibitemOpen
  \bibfield  {author} {\bibinfo {author} {\bibfnamefont {S.~P.}\ \bibnamefont
  {Martin}},\ }\bibfield  {title} {\bibinfo {title} {{Some simple criteria for
  gauged R-parity}},\ }\href {https://doi.org/10.1103/PhysRevD.46.R2769}
  {\bibfield  {journal} {\bibinfo  {journal} {Phys. Rev. D}\ }\textbf {\bibinfo
  {volume} {46}},\ \bibinfo {pages} {R2769} (\bibinfo {year} {1992})},\ \Eprint
  {https://arxiv.org/abs/hep-ph/9207218} {arXiv:hep-ph/9207218} \BibitemShut
  {NoStop}%
\bibitem [{\citenamefont {Harvey}\ and\ \citenamefont
  {Turner}(1990)}]{Harvey:1990qw}%
  \BibitemOpen
  \bibfield  {author} {\bibinfo {author} {\bibfnamefont {J.~A.}\ \bibnamefont
  {Harvey}}\ and\ \bibinfo {author} {\bibfnamefont {M.~S.}\ \bibnamefont
  {Turner}},\ }\bibfield  {title} {\bibinfo {title} {{Cosmological baryon and
  lepton number in the presence of electroweak fermion number violation}},\
  }\href {https://doi.org/10.1103/PhysRevD.42.3344} {\bibfield  {journal}
  {\bibinfo  {journal} {Phys. Rev. D}\ }\textbf {\bibinfo {volume} {42}},\
  \bibinfo {pages} {3344} (\bibinfo {year} {1990})}\BibitemShut {NoStop}%
\bibitem [{\citenamefont {Enqvist}\ \emph {et~al.}(2014)\citenamefont
  {Enqvist}, \citenamefont {Nurmi}, \citenamefont {Tenkanen},\ and\
  \citenamefont {Tuominen}}]{Enqvist:2014zqa}%
  \BibitemOpen
  \bibfield  {author} {\bibinfo {author} {\bibfnamefont {K.}~\bibnamefont
  {Enqvist}}, \bibinfo {author} {\bibfnamefont {S.}~\bibnamefont {Nurmi}},
  \bibinfo {author} {\bibfnamefont {T.}~\bibnamefont {Tenkanen}},\ and\
  \bibinfo {author} {\bibfnamefont {K.}~\bibnamefont {Tuominen}},\ }\bibfield
  {title} {\bibinfo {title} {{Standard Model with a real singlet scalar and
  inflation}},\ }\href {https://doi.org/10.1088/1475-7516/2014/08/035}
  {\bibfield  {journal} {\bibinfo  {journal} {JCAP}\ }\textbf {\bibinfo
  {volume} {08}},\ \bibinfo {pages} {035}},\ \Eprint
  {https://arxiv.org/abs/1407.0659} {arXiv:1407.0659 [astro-ph.CO]}
  \BibitemShut {NoStop}%
\bibitem [{\citenamefont {Cline}\ and\ \citenamefont
  {Kainulainen}(2013)}]{Cline:2012hg}%
  \BibitemOpen
  \bibfield  {author} {\bibinfo {author} {\bibfnamefont {J.~M.}\ \bibnamefont
  {Cline}}\ and\ \bibinfo {author} {\bibfnamefont {K.}~\bibnamefont
  {Kainulainen}},\ }\bibfield  {title} {\bibinfo {title} {{Electroweak
  baryogenesis and dark matter from a singlet Higgs}},\ }\href
  {https://doi.org/10.1088/1475-7516/2013/01/012} {\bibfield  {journal}
  {\bibinfo  {journal} {JCAP}\ }\textbf {\bibinfo {volume} {01}},\ \bibinfo
  {pages} {012}},\ \Eprint {https://arxiv.org/abs/1210.4196} {arXiv:1210.4196
  [hep-ph]} \BibitemShut {NoStop}%
\bibitem [{\citenamefont {Esteban}\ \emph {et~al.}(2024)\citenamefont
  {Esteban}, \citenamefont {Gonzalez-Garcia}, \citenamefont {Maltoni},
  \citenamefont {Martinez-Soler}, \citenamefont {Pinheiro},\ and\ \citenamefont
  {Schwetz}}]{esteban2024nufit60updatedglobalanalysis}%
  \BibitemOpen
  \bibfield  {author} {\bibinfo {author} {\bibfnamefont {I.}~\bibnamefont
  {Esteban}}, \bibinfo {author} {\bibfnamefont {M.~C.}\ \bibnamefont
  {Gonzalez-Garcia}}, \bibinfo {author} {\bibfnamefont {M.}~\bibnamefont
  {Maltoni}}, \bibinfo {author} {\bibfnamefont {I.}~\bibnamefont
  {Martinez-Soler}}, \bibinfo {author} {\bibfnamefont {J.~P.}\ \bibnamefont
  {Pinheiro}},\ and\ \bibinfo {author} {\bibfnamefont {T.}~\bibnamefont
  {Schwetz}},\ }\href {https://arxiv.org/abs/2410.05380} {\bibinfo {title}
  {Nufit-6.0: Updated global analysis of three-flavor neutrino oscillations}}
  (\bibinfo {year} {2024}),\ \Eprint {https://arxiv.org/abs/2410.05380}
  {arXiv:2410.05380 [hep-ph]} \BibitemShut {NoStop}%
\bibitem [{\citenamefont {Herrero-Garcia}\ \emph {et~al.}(2018)\citenamefont
  {Herrero-Garcia}, \citenamefont {Molinaro},\ and\ \citenamefont
  {Schmidt}}]{Herrero-Garcia:2018koq}%
  \BibitemOpen
  \bibfield  {author} {\bibinfo {author} {\bibfnamefont {J.}~\bibnamefont
  {Herrero-Garcia}}, \bibinfo {author} {\bibfnamefont {E.}~\bibnamefont
  {Molinaro}},\ and\ \bibinfo {author} {\bibfnamefont {M.~A.}\ \bibnamefont
  {Schmidt}},\ }\bibfield  {title} {\bibinfo {title} {{Dark matter direct
  detection of a fermionic singlet at one loop}},\ }\href
  {https://doi.org/10.1140/epjc/s10052-018-5935-5} {\bibfield  {journal}
  {\bibinfo  {journal} {Eur. Phys. J. C}\ }\textbf {\bibinfo {volume} {78}},\
  \bibinfo {pages} {471} (\bibinfo {year} {2018})},\ \bibinfo {note} {[Erratum:
  None 82, 53 (2022)]},\ \Eprint {https://arxiv.org/abs/1803.05660}
  {arXiv:1803.05660 [hep-ph]} \BibitemShut {NoStop}%
\bibitem [{\citenamefont {Ir\v{s}i\v{c}}\ \emph {et~al.}(2017)\citenamefont
  {Ir\v{s}i\v{c}} \emph {et~al.}}]{Irsic:2017ixq}%
  \BibitemOpen
  \bibfield  {author} {\bibinfo {author} {\bibfnamefont {V.}~\bibnamefont
  {Ir\v{s}i\v{c}}} \emph {et~al.},\ }\bibfield  {title} {\bibinfo {title} {{New
  Constraints on the free-streaming of warm dark matter from intermediate and
  small scale Lyman-$\alpha$ forest data}},\ }\href
  {https://doi.org/10.1103/PhysRevD.96.023522} {\bibfield  {journal} {\bibinfo
  {journal} {Phys. Rev. D}\ }\textbf {\bibinfo {volume} {96}},\ \bibinfo
  {pages} {023522} (\bibinfo {year} {2017})},\ \Eprint
  {https://arxiv.org/abs/1702.01764} {arXiv:1702.01764 [astro-ph.CO]}
  \BibitemShut {NoStop}%
\bibitem [{\citenamefont {Boyarsky}\ \emph {et~al.}(2009)\citenamefont
  {Boyarsky}, \citenamefont {Lesgourgues}, \citenamefont {Ruchayskiy},\ and\
  \citenamefont {Viel}}]{Boyarsky:2008xj}%
  \BibitemOpen
  \bibfield  {author} {\bibinfo {author} {\bibfnamefont {A.}~\bibnamefont
  {Boyarsky}}, \bibinfo {author} {\bibfnamefont {J.}~\bibnamefont
  {Lesgourgues}}, \bibinfo {author} {\bibfnamefont {O.}~\bibnamefont
  {Ruchayskiy}},\ and\ \bibinfo {author} {\bibfnamefont {M.}~\bibnamefont
  {Viel}},\ }\bibfield  {title} {\bibinfo {title} {{Lyman-alpha constraints on
  warm and on warm-plus-cold dark matter models}},\ }\href
  {https://doi.org/10.1088/1475-7516/2009/05/012} {\bibfield  {journal}
  {\bibinfo  {journal} {JCAP}\ }\textbf {\bibinfo {volume} {05}},\ \bibinfo
  {pages} {012}},\ \Eprint {https://arxiv.org/abs/0812.0010} {arXiv:0812.0010
  [astro-ph]} \BibitemShut {NoStop}%
\bibitem [{\citenamefont {Decant}\ \emph {et~al.}(2022)\citenamefont {Decant},
  \citenamefont {Heisig}, \citenamefont {Hooper},\ and\ \citenamefont
  {Lopez-Honorez}}]{Decant:2021mhj}%
  \BibitemOpen
  \bibfield  {author} {\bibinfo {author} {\bibfnamefont {Q.}~\bibnamefont
  {Decant}}, \bibinfo {author} {\bibfnamefont {J.}~\bibnamefont {Heisig}},
  \bibinfo {author} {\bibfnamefont {D.~C.}\ \bibnamefont {Hooper}},\ and\
  \bibinfo {author} {\bibfnamefont {L.}~\bibnamefont {Lopez-Honorez}},\
  }\bibfield  {title} {\bibinfo {title} {{Lyman-\ensuremath{\alpha} constraints
  on freeze-in and superWIMPs}},\ }\href
  {https://doi.org/10.1088/1475-7516/2022/03/041} {\bibfield  {journal}
  {\bibinfo  {journal} {JCAP}\ }\textbf {\bibinfo {volume} {03}},\ \bibinfo
  {pages} {041}},\ \Eprint {https://arxiv.org/abs/2111.09321} {arXiv:2111.09321
  [astro-ph.CO]} \BibitemShut {NoStop}%
\bibitem [{\citenamefont {Cirelli}\ \emph {et~al.}(2024)\citenamefont
  {Cirelli}, \citenamefont {Strumia},\ and\ \citenamefont
  {Zupan}}]{Cirelli:2024ssz}%
  \BibitemOpen
  \bibfield  {author} {\bibinfo {author} {\bibfnamefont {M.}~\bibnamefont
  {Cirelli}}, \bibinfo {author} {\bibfnamefont {A.}~\bibnamefont {Strumia}},\
  and\ \bibinfo {author} {\bibfnamefont {J.}~\bibnamefont {Zupan}},\ }\bibfield
   {title} {\bibinfo {title} {{Dark Matter}},\ }\href@noop {} {\  (\bibinfo
  {year} {2024})},\ \Eprint {https://arxiv.org/abs/2406.01705}
  {arXiv:2406.01705 [hep-ph]} \BibitemShut {NoStop}%
\bibitem [{\citenamefont {Ade}\ \emph {et~al.}(2019)\citenamefont {Ade} \emph
  {et~al.}}]{SimonsObservatory:2018koc}%
  \BibitemOpen
  \bibfield  {author} {\bibinfo {author} {\bibfnamefont {P.}~\bibnamefont
  {Ade}} \emph {et~al.} (\bibinfo {collaboration} {Simons Observatory}),\
  }\bibfield  {title} {\bibinfo {title} {{The Simons Observatory: Science goals
  and forecasts}},\ }\href {https://doi.org/10.1088/1475-7516/2019/02/056}
  {\bibfield  {journal} {\bibinfo  {journal} {JCAP}\ }\textbf {\bibinfo
  {volume} {02}},\ \bibinfo {pages} {056}},\ \Eprint
  {https://arxiv.org/abs/1808.07445} {arXiv:1808.07445 [astro-ph.CO]}
  \BibitemShut {NoStop}%
\bibitem [{\citenamefont {Laureijs}\ \emph {et~al.}(2011)\citenamefont
  {Laureijs} \emph {et~al.}}]{EUCLID:2011zbd}%
  \BibitemOpen
  \bibfield  {author} {\bibinfo {author} {\bibfnamefont {R.}~\bibnamefont
  {Laureijs}} \emph {et~al.} (\bibinfo {collaboration} {EUCLID}),\ }\bibfield
  {title} {\bibinfo {title} {{Euclid Definition Study Report}},\ }\href@noop {}
  {\  (\bibinfo {year} {2011})},\ \Eprint {https://arxiv.org/abs/1110.3193}
  {arXiv:1110.3193 [astro-ph.CO]} \BibitemShut {NoStop}%
\bibitem [{\citenamefont {Aiola}\ \emph {et~al.}(2022)\citenamefont {Aiola}
  \emph {et~al.}}]{CMB-HD:2022bsz}%
  \BibitemOpen
  \bibfield  {author} {\bibinfo {author} {\bibfnamefont {S.}~\bibnamefont
  {Aiola}} \emph {et~al.} (\bibinfo {collaboration} {CMB-HD}),\ }\bibfield
  {title} {\bibinfo {title} {{Snowmass2021 CMB-HD White Paper}},\ }\href@noop
  {} {\  (\bibinfo {year} {2022})},\ \Eprint {https://arxiv.org/abs/2203.05728}
  {arXiv:2203.05728 [astro-ph.CO]} \BibitemShut {NoStop}%
\bibitem [{\citenamefont {Kanzaki}\ \emph {et~al.}(2007)\citenamefont
  {Kanzaki}, \citenamefont {Kawasaki}, \citenamefont {Kohri},\ and\
  \citenamefont {Moroi}}]{Kanzaki:2007pd}%
  \BibitemOpen
  \bibfield  {author} {\bibinfo {author} {\bibfnamefont {T.}~\bibnamefont
  {Kanzaki}}, \bibinfo {author} {\bibfnamefont {M.}~\bibnamefont {Kawasaki}},
  \bibinfo {author} {\bibfnamefont {K.}~\bibnamefont {Kohri}},\ and\ \bibinfo
  {author} {\bibfnamefont {T.}~\bibnamefont {Moroi}},\ }\bibfield  {title}
  {\bibinfo {title} {{Cosmological Constraints on Neutrino Injection}},\ }\href
  {https://doi.org/10.1103/PhysRevD.76.105017} {\bibfield  {journal} {\bibinfo
  {journal} {Phys. Rev. D}\ }\textbf {\bibinfo {volume} {76}},\ \bibinfo
  {pages} {105017} (\bibinfo {year} {2007})},\ \Eprint
  {https://arxiv.org/abs/0705.1200} {arXiv:0705.1200 [hep-ph]} \BibitemShut
  {NoStop}%
\bibitem [{\citenamefont {Poulin}\ \emph {et~al.}(2017)\citenamefont {Poulin},
  \citenamefont {Lesgourgues},\ and\ \citenamefont {Serpico}}]{Poulin:2016anj}%
  \BibitemOpen
  \bibfield  {author} {\bibinfo {author} {\bibfnamefont {V.}~\bibnamefont
  {Poulin}}, \bibinfo {author} {\bibfnamefont {J.}~\bibnamefont
  {Lesgourgues}},\ and\ \bibinfo {author} {\bibfnamefont {P.~D.}\ \bibnamefont
  {Serpico}},\ }\bibfield  {title} {\bibinfo {title} {{Cosmological constraints
  on exotic injection of electromagnetic energy}},\ }\href
  {https://doi.org/10.1088/1475-7516/2017/03/043} {\bibfield  {journal}
  {\bibinfo  {journal} {JCAP}\ }\textbf {\bibinfo {volume} {03}},\ \bibinfo
  {pages} {043}},\ \Eprint {https://arxiv.org/abs/1610.10051} {arXiv:1610.10051
  [astro-ph.CO]} \BibitemShut {NoStop}%
\bibitem [{\citenamefont {Chluba}\ \emph {et~al.}(2019)\citenamefont {Chluba}
  \emph {et~al.}}]{Chluba:2019kpb}%
  \BibitemOpen
  \bibfield  {author} {\bibinfo {author} {\bibfnamefont {J.}~\bibnamefont
  {Chluba}} \emph {et~al.},\ }\bibfield  {title} {\bibinfo {title} {{Spectral
  Distortions of the CMB as a Probe of Inflation, Recombination, Structure
  Formation and Particle Physics}: {Astro2020 Science White Paper}},\
  }\href@noop {} {\bibfield  {journal} {\bibinfo  {journal} {Bull. Am. Astron.
  Soc.}\ }\textbf {\bibinfo {volume} {51}},\ \bibinfo {pages} {184} (\bibinfo
  {year} {2019})},\ \Eprint {https://arxiv.org/abs/1903.04218}
  {arXiv:1903.04218 [astro-ph.CO]} \BibitemShut {NoStop}%
\bibitem [{\citenamefont {Bauer}\ \emph {et~al.}(2021)\citenamefont {Bauer},
  \citenamefont {Rodd},\ and\ \citenamefont {Webber}}]{Bauer:2020jay}%
  \BibitemOpen
  \bibfield  {author} {\bibinfo {author} {\bibfnamefont {C.~W.}\ \bibnamefont
  {Bauer}}, \bibinfo {author} {\bibfnamefont {N.~L.}\ \bibnamefont {Rodd}},\
  and\ \bibinfo {author} {\bibfnamefont {B.~R.}\ \bibnamefont {Webber}},\
  }\bibfield  {title} {\bibinfo {title} {{Dark matter spectra from the
  electroweak to the Planck scale}},\ }\href
  {https://doi.org/10.1007/JHEP06(2021)121} {\bibfield  {journal} {\bibinfo
  {journal} {JHEP}\ }\textbf {\bibinfo {volume} {06}},\ \bibinfo {pages}
  {121}},\ \Eprint {https://arxiv.org/abs/2007.15001} {arXiv:2007.15001
  [hep-ph]} \BibitemShut {NoStop}%
\bibitem [{\citenamefont {Chluba}\ \emph {et~al.}(2020)\citenamefont {Chluba},
  \citenamefont {Ravenni},\ and\ \citenamefont {Acharya}}]{Chluba_2020}%
  \BibitemOpen
  \bibfield  {author} {\bibinfo {author} {\bibfnamefont {J.}~\bibnamefont
  {Chluba}}, \bibinfo {author} {\bibfnamefont {A.}~\bibnamefont {Ravenni}},\
  and\ \bibinfo {author} {\bibfnamefont {S.~K.}\ \bibnamefont {Acharya}},\
  }\bibfield  {title} {\bibinfo {title} {Thermalization of large energy release
  in the early universe},\ }\href {https://doi.org/10.1093/mnras/staa2131}
  {\bibfield  {journal} {\bibinfo  {journal} {Monthly Notices of the Royal
  Astronomical Society}\ }\textbf {\bibinfo {volume} {498}},\ \bibinfo {pages}
  {959–980} (\bibinfo {year} {2020})}\BibitemShut {NoStop}%
\bibitem [{\citenamefont {Mather}\ \emph {et~al.}(1994)\citenamefont {Mather}
  \emph {et~al.}}]{Mather:1993ij}%
  \BibitemOpen
  \bibfield  {author} {\bibinfo {author} {\bibfnamefont {J.~C.}\ \bibnamefont
  {Mather}} \emph {et~al.},\ }\bibfield  {title} {\bibinfo {title}
  {{Measurement of the Cosmic Microwave Background spectrum by the COBE FIRAS
  instrument}},\ }\href {https://doi.org/10.1086/173574} {\bibfield  {journal}
  {\bibinfo  {journal} {Astrophys. J.}\ }\textbf {\bibinfo {volume} {420}},\
  \bibinfo {pages} {439} (\bibinfo {year} {1994})}\BibitemShut {NoStop}%
\bibitem [{\citenamefont {Fixsen}\ \emph {et~al.}(1996)\citenamefont {Fixsen},
  \citenamefont {Cheng}, \citenamefont {Gales}, \citenamefont {Mather},
  \citenamefont {Shafer},\ and\ \citenamefont {Wright}}]{Fixsen:1996nj}%
  \BibitemOpen
  \bibfield  {author} {\bibinfo {author} {\bibfnamefont {D.~J.}\ \bibnamefont
  {Fixsen}}, \bibinfo {author} {\bibfnamefont {E.~S.}\ \bibnamefont {Cheng}},
  \bibinfo {author} {\bibfnamefont {J.~M.}\ \bibnamefont {Gales}}, \bibinfo
  {author} {\bibfnamefont {J.~C.}\ \bibnamefont {Mather}}, \bibinfo {author}
  {\bibfnamefont {R.~A.}\ \bibnamefont {Shafer}},\ and\ \bibinfo {author}
  {\bibfnamefont {E.~L.}\ \bibnamefont {Wright}},\ }\bibfield  {title}
  {\bibinfo {title} {{The Cosmic Microwave Background spectrum from the full
  COBE FIRAS data set}},\ }\href {https://doi.org/10.1086/178173} {\bibfield
  {journal} {\bibinfo  {journal} {Astrophys. J.}\ }\textbf {\bibinfo {volume}
  {473}},\ \bibinfo {pages} {576} (\bibinfo {year} {1996})},\ \Eprint
  {https://arxiv.org/abs/astro-ph/9605054} {arXiv:astro-ph/9605054}
  \BibitemShut {NoStop}%
\bibitem [{\citenamefont {Li}\ \emph {et~al.}(2021)\citenamefont {Li},
  \citenamefont {Li}, \citenamefont {Yan},\ and\ \citenamefont
  {Yang}}]{Li_2021}%
  \BibitemOpen
  \bibfield  {author} {\bibinfo {author} {\bibfnamefont {S.-P.}\ \bibnamefont
  {Li}}, \bibinfo {author} {\bibfnamefont {X.-Q.}\ \bibnamefont {Li}}, \bibinfo
  {author} {\bibfnamefont {X.-S.}\ \bibnamefont {Yan}},\ and\ \bibinfo {author}
  {\bibfnamefont {Y.-D.}\ \bibnamefont {Yang}},\ }\bibfield  {title} {\bibinfo
  {title} {Simple estimate of bbn sensitivity to light freeze-in dark matter},\
  }\bibfield  {journal} {\bibinfo  {journal} {Physical Review D}\ }\textbf
  {\bibinfo {volume} {104}},\ \href
  {https://doi.org/10.1103/physrevd.104.115007} {10.1103/physrevd.104.115007}
  (\bibinfo {year} {2021})\BibitemShut {NoStop}%
\bibitem [{\citenamefont {Kogut}\ \emph {et~al.}(2011)\citenamefont {Kogut}
  \emph {et~al.}}]{Kogut:2011xw}%
  \BibitemOpen
  \bibfield  {author} {\bibinfo {author} {\bibfnamefont {A.}~\bibnamefont
  {Kogut}} \emph {et~al.},\ }\bibfield  {title} {\bibinfo {title} {{The
  Primordial Inflation Explorer (PIXIE): A Nulling Polarimeter for Cosmic
  Microwave Background Observations}},\ }\href
  {https://doi.org/10.1088/1475-7516/2011/07/025} {\bibfield  {journal}
  {\bibinfo  {journal} {JCAP}\ }\textbf {\bibinfo {volume} {07}},\ \bibinfo
  {pages} {025}},\ \Eprint {https://arxiv.org/abs/1105.2044} {arXiv:1105.2044
  [astro-ph.CO]} \BibitemShut {NoStop}%
\bibitem [{\citenamefont {Feldstein}\ \emph {et~al.}(2013)\citenamefont
  {Feldstein}, \citenamefont {Kusenko}, \citenamefont {Matsumoto},\ and\
  \citenamefont {Yanagida}}]{Feldstein:2013kka}%
  \BibitemOpen
  \bibfield  {author} {\bibinfo {author} {\bibfnamefont {B.}~\bibnamefont
  {Feldstein}}, \bibinfo {author} {\bibfnamefont {A.}~\bibnamefont {Kusenko}},
  \bibinfo {author} {\bibfnamefont {S.}~\bibnamefont {Matsumoto}},\ and\
  \bibinfo {author} {\bibfnamefont {T.~T.}\ \bibnamefont {Yanagida}},\
  }\bibfield  {title} {\bibinfo {title} {{Neutrinos at IceCube from Heavy
  Decaying Dark Matter}},\ }\href {https://doi.org/10.1103/PhysRevD.88.015004}
  {\bibfield  {journal} {\bibinfo  {journal} {Phys. Rev. D}\ }\textbf {\bibinfo
  {volume} {88}},\ \bibinfo {pages} {015004} (\bibinfo {year} {2013})},\
  \Eprint {https://arxiv.org/abs/1303.7320} {arXiv:1303.7320 [hep-ph]}
  \BibitemShut {NoStop}%
\bibitem [{\citenamefont {Palomares-Ruiz}(2008)}]{Palomares-Ruiz:2007egs}%
  \BibitemOpen
  \bibfield  {author} {\bibinfo {author} {\bibfnamefont {S.}~\bibnamefont
  {Palomares-Ruiz}},\ }\bibfield  {title} {\bibinfo {title} {{Model-independent
  bound on the dark matter lifetime}},\ }\href
  {https://doi.org/10.1016/j.physletb.2008.05.040} {\bibfield  {journal}
  {\bibinfo  {journal} {Phys. Lett. B}\ }\textbf {\bibinfo {volume} {665}},\
  \bibinfo {pages} {50} (\bibinfo {year} {2008})},\ \Eprint
  {https://arxiv.org/abs/0712.1937} {arXiv:0712.1937 [astro-ph]} \BibitemShut
  {NoStop}%
\bibitem [{\citenamefont {Bell}\ \emph {et~al.}(2010)\citenamefont {Bell},
  \citenamefont {Galea},\ and\ \citenamefont {Petraki}}]{Bell:2010fk}%
  \BibitemOpen
  \bibfield  {author} {\bibinfo {author} {\bibfnamefont {N.~F.}\ \bibnamefont
  {Bell}}, \bibinfo {author} {\bibfnamefont {A.~J.}\ \bibnamefont {Galea}},\
  and\ \bibinfo {author} {\bibfnamefont {K.}~\bibnamefont {Petraki}},\
  }\bibfield  {title} {\bibinfo {title} {{Lifetime Constraints for Late Dark
  Matter Decay}},\ }\href {https://doi.org/10.1103/PhysRevD.82.023514}
  {\bibfield  {journal} {\bibinfo  {journal} {Phys. Rev. D}\ }\textbf {\bibinfo
  {volume} {82}},\ \bibinfo {pages} {023514} (\bibinfo {year} {2010})},\
  \Eprint {https://arxiv.org/abs/1004.1008} {arXiv:1004.1008 [astro-ph.HE]}
  \BibitemShut {NoStop}%
\bibitem [{\citenamefont {Esmaili}\ \emph {et~al.}(2012)\citenamefont
  {Esmaili}, \citenamefont {Ibarra},\ and\ \citenamefont
  {Peres}}]{Esmaili:2012us}%
  \BibitemOpen
  \bibfield  {author} {\bibinfo {author} {\bibfnamefont {A.}~\bibnamefont
  {Esmaili}}, \bibinfo {author} {\bibfnamefont {A.}~\bibnamefont {Ibarra}},\
  and\ \bibinfo {author} {\bibfnamefont {O.~L.~G.}\ \bibnamefont {Peres}},\
  }\bibfield  {title} {\bibinfo {title} {{Probing the stability of superheavy
  dark matter particles with high-energy neutrinos}},\ }\href
  {https://doi.org/10.1088/1475-7516/2012/11/034} {\bibfield  {journal}
  {\bibinfo  {journal} {JCAP}\ }\textbf {\bibinfo {volume} {11}},\ \bibinfo
  {pages} {034}},\ \Eprint {https://arxiv.org/abs/1205.5281} {arXiv:1205.5281
  [hep-ph]} \BibitemShut {NoStop}%
\bibitem [{\citenamefont {Poulin}\ \emph {et~al.}(2016)\citenamefont {Poulin},
  \citenamefont {Serpico},\ and\ \citenamefont {Lesgourgues}}]{Poulin:2016nat}%
  \BibitemOpen
  \bibfield  {author} {\bibinfo {author} {\bibfnamefont {V.}~\bibnamefont
  {Poulin}}, \bibinfo {author} {\bibfnamefont {P.~D.}\ \bibnamefont
  {Serpico}},\ and\ \bibinfo {author} {\bibfnamefont {J.}~\bibnamefont
  {Lesgourgues}},\ }\bibfield  {title} {\bibinfo {title} {{A fresh look at
  linear cosmological constraints on a decaying dark matter component}},\
  }\href {https://doi.org/10.1088/1475-7516/2016/08/036} {\bibfield  {journal}
  {\bibinfo  {journal} {JCAP}\ }\textbf {\bibinfo {volume} {08}},\ \bibinfo
  {pages} {036}},\ \Eprint {https://arxiv.org/abs/1606.02073} {arXiv:1606.02073
  [astro-ph.CO]} \BibitemShut {NoStop}%
\bibitem [{\citenamefont {Arg\"uelles}\ \emph {et~al.}(2021)\citenamefont
  {Arg\"uelles}, \citenamefont {Diaz}, \citenamefont {Kheirandish},
  \citenamefont {Olivares-Del-Campo}, \citenamefont {Safa},\ and\ \citenamefont
  {Vincent}}]{Arguelles:2019ouk}%
  \BibitemOpen
  \bibfield  {author} {\bibinfo {author} {\bibfnamefont {C.~A.}\ \bibnamefont
  {Arg\"uelles}}, \bibinfo {author} {\bibfnamefont {A.}~\bibnamefont {Diaz}},
  \bibinfo {author} {\bibfnamefont {A.}~\bibnamefont {Kheirandish}}, \bibinfo
  {author} {\bibfnamefont {A.}~\bibnamefont {Olivares-Del-Campo}}, \bibinfo
  {author} {\bibfnamefont {I.}~\bibnamefont {Safa}},\ and\ \bibinfo {author}
  {\bibfnamefont {A.~C.}\ \bibnamefont {Vincent}},\ }\bibfield  {title}
  {\bibinfo {title} {{Dark matter annihilation to neutrinos}},\ }\href
  {https://doi.org/10.1103/RevModPhys.93.035007} {\bibfield  {journal}
  {\bibinfo  {journal} {Rev. Mod. Phys.}\ }\textbf {\bibinfo {volume} {93}},\
  \bibinfo {pages} {035007} (\bibinfo {year} {2021})},\ \Eprint
  {https://arxiv.org/abs/1912.09486} {arXiv:1912.09486 [hep-ph]} \BibitemShut
  {NoStop}%
\bibitem [{\citenamefont {Palomares-Ruiz}(2020)}]{Palomares-Ruiz:2020ytu}%
  \BibitemOpen
  \bibfield  {author} {\bibinfo {author} {\bibfnamefont {S.}~\bibnamefont
  {Palomares-Ruiz}},\ }\bibinfo {title} {{Tests of Dark Matter Scenarios with
  Neutrino Telescopes}},\ in\ \href
  {https://doi.org/10.1142/9789813275027_0007} {\emph {\bibinfo {booktitle}
  {{Probing Particle Physics with Neutrino Telescopes}}}}\ (\bibinfo {year}
  {2020})\ pp.\ \bibinfo {pages} {191--266}\BibitemShut {NoStop}%
\bibitem [{\citenamefont {Fukuda}\ \emph {et~al.}(2003)\citenamefont {Fukuda}
  \emph {et~al.}}]{Super-Kamiokande:2002weg}%
  \BibitemOpen
  \bibfield  {author} {\bibinfo {author} {\bibfnamefont {Y.}~\bibnamefont
  {Fukuda}} \emph {et~al.} (\bibinfo {collaboration} {Super-Kamiokande}),\
  }\bibfield  {title} {\bibinfo {title} {{The Super-Kamiokande detector}},\
  }\href {https://doi.org/10.1016/S0168-9002(03)00425-X} {\bibfield  {journal}
  {\bibinfo  {journal} {Nucl. Instrum. Meth. A}\ }\textbf {\bibinfo {volume}
  {501}},\ \bibinfo {pages} {418} (\bibinfo {year} {2003})}\BibitemShut
  {NoStop}%
\bibitem [{\citenamefont {Abe}\ \emph {et~al.}(2018)\citenamefont {Abe} \emph
  {et~al.}}]{Hyper-Kamiokande:2018ofw}%
  \BibitemOpen
  \bibfield  {author} {\bibinfo {author} {\bibfnamefont {K.}~\bibnamefont
  {Abe}} \emph {et~al.} (\bibinfo {collaboration} {Hyper-Kamiokande}),\
  }\bibfield  {title} {\bibinfo {title} {{Hyper-Kamiokande Design Report}},\
  }\href@noop {} {\  (\bibinfo {year} {2018})},\ \Eprint
  {https://arxiv.org/abs/1805.04163} {arXiv:1805.04163 [physics.ins-det]}
  \BibitemShut {NoStop}%
\bibitem [{\citenamefont {et~al}(2013)}]{Aartsen_2013}%
  \BibitemOpen
  \bibfield  {author} {\bibinfo {author} {\bibfnamefont {A.}~\bibnamefont
  {et~al}},\ }\bibfield  {title} {\bibinfo {title} {First observation of
  pev-energy neutrinos with icecube},\ }\bibfield  {journal} {\bibinfo
  {journal} {Physical Review Letters}\ }\textbf {\bibinfo {volume} {111}},\
  \href {https://doi.org/10.1103/physrevlett.111.021103}
  {10.1103/physrevlett.111.021103} (\bibinfo {year} {2013})\BibitemShut
  {NoStop}%
\bibitem [{\citenamefont {Aartsen}\ \emph {et~al.}(2018)\citenamefont {Aartsen}
  \emph {et~al.}}]{IceCube:2018tkk}%
  \BibitemOpen
  \bibfield  {author} {\bibinfo {author} {\bibfnamefont {M.~G.}\ \bibnamefont
  {Aartsen}} \emph {et~al.} (\bibinfo {collaboration} {IceCube}),\ }\bibfield
  {title} {\bibinfo {title} {{Search for neutrinos from decaying dark matter
  with IceCube}},\ }\href {https://doi.org/10.1140/epjc/s10052-018-6273-3}
  {\bibfield  {journal} {\bibinfo  {journal} {Eur. Phys. J. C}\ }\textbf
  {\bibinfo {volume} {78}},\ \bibinfo {pages} {831} (\bibinfo {year} {2018})},\
  \Eprint {https://arxiv.org/abs/1804.03848} {arXiv:1804.03848 [astro-ph.HE]}
  \BibitemShut {NoStop}%
\bibitem [{\citenamefont {Adrian-Martinez}\ \emph {et~al.}(2015)\citenamefont
  {Adrian-Martinez} \emph {et~al.}}]{ANTARES:2015vis}%
  \BibitemOpen
  \bibfield  {author} {\bibinfo {author} {\bibfnamefont {S.}~\bibnamefont
  {Adrian-Martinez}} \emph {et~al.} (\bibinfo {collaboration} {ANTARES}),\
  }\bibfield  {title} {\bibinfo {title} {{Search of Dark Matter Annihilation in
  the Galactic Centre using the ANTARES Neutrino Telescope}},\ }\href
  {https://doi.org/10.1088/1475-7516/2015/10/068} {\bibfield  {journal}
  {\bibinfo  {journal} {JCAP}\ }\textbf {\bibinfo {volume} {10}},\ \bibinfo
  {pages} {068}},\ \Eprint {https://arxiv.org/abs/1505.04866} {arXiv:1505.04866
  [astro-ph.HE]} \BibitemShut {NoStop}%
\bibitem [{\citenamefont {Adrian-Martinez}\ \emph {et~al.}(2016)\citenamefont
  {Adrian-Martinez} \emph {et~al.}}]{KM3Net:2016zxf}%
  \BibitemOpen
  \bibfield  {author} {\bibinfo {author} {\bibfnamefont {S.}~\bibnamefont
  {Adrian-Martinez}} \emph {et~al.} (\bibinfo {collaboration} {KM3Net}),\
  }\bibfield  {title} {\bibinfo {title} {{Letter of intent for KM3NeT 2.0}},\
  }\href {https://doi.org/10.1088/0954-3899/43/8/084001} {\bibfield  {journal}
  {\bibinfo  {journal} {J. Phys. G}\ }\textbf {\bibinfo {volume} {43}},\
  \bibinfo {pages} {084001} (\bibinfo {year} {2016})},\ \Eprint
  {https://arxiv.org/abs/1601.07459} {arXiv:1601.07459 [astro-ph.IM]}
  \BibitemShut {NoStop}%
\bibitem [{\citenamefont {Aiello}\ \emph {et~al.}(2019)\citenamefont {Aiello}
  \emph {et~al.}}]{KM3NeT:2018wnd}%
  \BibitemOpen
  \bibfield  {author} {\bibinfo {author} {\bibfnamefont {S.}~\bibnamefont
  {Aiello}} \emph {et~al.} (\bibinfo {collaboration} {KM3NeT}),\ }\bibfield
  {title} {\bibinfo {title} {{Sensitivity of the KM3NeT/ARCA neutrino telescope
  to point-like neutrino sources}},\ }\href
  {https://doi.org/10.1016/j.astropartphys.2019.04.002} {\bibfield  {journal}
  {\bibinfo  {journal} {Astropart. Phys.}\ }\textbf {\bibinfo {volume} {111}},\
  \bibinfo {pages} {100} (\bibinfo {year} {2019})},\ \Eprint
  {https://arxiv.org/abs/1810.08499} {arXiv:1810.08499 [astro-ph.HE]}
  \BibitemShut {NoStop}%
\end{thebibliography}%

\end{document}